\documentclass[final]{IEEEtran}

\usepackage{amsthm,amssymb,graphicx,multirow,amsmath,color,amsfonts,mathtools}
\usepackage[update,prepend]{epstopdf}
\usepackage[noadjust]{cite}
\usepackage[latin1]{inputenc}
\usepackage{tikz}
\usetikzlibrary{arrows,calc} 
\usepackage{bbm} 
\usepackage{pdfpages}
\usepackage{tabulary}
\usepackage{multirow}
\usepackage{comment}
\usepackage{subfigure}

\usepackage[T1]{fontenc}
\usepackage[font=small]{caption}

\def\nb0{{\mathbf{0}}}
\def\nb1{{\mathbf{1}}}

\def\ncalF{{\mathcal{F}}}

\def\ncalN{{\mathcal{N}}}

\def\nbbE{{\mathbb{E}}}

\def\sinc{{\rm sinc}}

\newtheorem{lemma}{Lemma}

\newtheorem{theorem}{Theorem}

\newtheorem{cor}{Corollary}

\newtheorem{remark}{Remark}

\allowdisplaybreaks 

\begin{document}
\graphicspath{{./Figures/}}
\title{\huge Fundamentals of Wobbling and Hardware Impairments-Aware Air-to-Ground Channel Model
}
\vspace{-0.2cm}
\author{
Morteza Banagar, \textit{Member, IEEE} and Harpreet S. Dhillon, \textit{Fellow, IEEE}
\thanks{The authors are with Wireless@VT, Department of ECE, Virginia Tech, Blacksburg, VA (email: \{mbanagar, hdhillon\}@vt.edu). The support of the US NSF (Grant CNS-1923807) is gratefully acknowledged. This article was presented in part at the 2022 IEEE Globecom in Rio de Janeiro, Brazil \cite{C_Morteza_Impairment_2022}.}
}

\maketitle
\vspace{-1.8cm}
\begin{abstract}
In this paper, we develop an impairments-aware air-to-ground unified channel model that incorporates the effect of both wobbling and hardware impairments, where the former is caused by random physical fluctuations of unmanned aerial vehicles (UAVs), and the latter by intrinsic radio frequency (RF) nonidealities at both the transmitter and receiver, such as phase noise, in-phase/quadrature (I/Q) imbalance, and power amplifier (PA) nonlinearity. The impact of UAV wobbling is modeled by two stochastic processes, i.e., the canonical Wiener process and the more realistic sinusoidal process. On the other hand, the \textit{aggregate} impact of all hardware impairments is modeled as two multiplicative and additive distortion noise processes, which is a well-accepted model. For the sake of generality, we consider both wide-sense stationary (WSS) and nonstationary processes for the distortion noises. We then rigorously characterize the autocorrelation function (ACF) of the wireless channel, using which we provide a comprehensive analysis of four key channel-related metrics: (i) power delay profile (PDP), (ii) coherence time, (iii) coherence bandwidth, and (iv) power spectral density (PSD) of the distortion-plus-noise process. Furthermore, we evaluate these metrics with reasonable UAV wobbling and hardware impairment models to obtain useful insights. Quite noticeably, we demonstrate that even for small UAV wobbling, the coherence time severely degrades at high frequencies, which renders air-to-ground channel estimation very difficult at these frequencies. To the best of our understanding, this is the first work that characterizes the joint impact of UAV wobbling and hardware impairments on the air-to-ground wireless channel.
\end{abstract}

\section{Introduction} \label{sec:Intro}
Owing to their agility, cost-effectiveness, ease of deployment, and higher probability of line-of-sight (LoS), UAVs have become an integral part of the current wireless ecosystem \cite{J_Mozaffari_Tutorial_2018, J_Zeng_Accessing_2019, D_Morteza_Drone_2022, J_Morteza_Performance_2019, B_Morteza_Stochastic_2020}. Given their proposed use in a variety of applications in different frequency bands, it is important to understand the fundamental characteristics of the air-to-ground wireless channels, which is a topic of significant current interest. To that end, our specific focus in this paper is on studying the joint impact of a unique UAV impairment (wobbling) and a fundamental characteristic of all transceivers (hardware impairments) on the properties of the air-to-ground wireless channels.

Because of the lack of fixed infrastructures, UAVs suffer from some amount of physical vibrations, especially under inclement weather conditions or high vibration frequency of their propellers \cite{C_Li_Development_2017, J_Morteza_Impact_2020}. These physical vibrations (also known as wobbling \cite{J_Morteza_Impact_2020}, jittering \cite{J_Xu_Multiuser_2020, J_Wu_Energy_2020}, and fluctuations \cite{J_Dabiri_Analytical_2020}) could adversely affect the wireless channel especially at high frequencies. Further, all digital transceivers suffer from RF nonidealities (also known as hardware impairments) that degrade or even severely limit their performance \cite{B_Schenk_Imperfections_2008, J_Emil_Massive_2014}. While the hardware impairments are well-understood in the literature, their interplay with the unique UAV impairments, such as wobbling, has not been studied yet. Motivated by this observation and viewing the UAV wobbling as an ``impairment", we consider an aerial-terrestrial setup, where the terrestrial node suffers from hardware impairments, while the aerial node suffers from both hardware impairments and wobbling. For this setup, we develop a unified air-to-ground channel model that is cognizant of these impairments and provide a comprehensive analysis of key channel-related metrics. We also discuss the implications of our analysis on operating UAVs at high frequencies. 

\subsection{Related Works} \label{subsec:Related}
This paper builds on the following general lines of research: (i) air-to-ground channel model, (ii) UAV wobbling, and (iii) hardware impairments. We explain each research direction next.

\textit{Air-to-Ground Channel Model.}
There have been numerous works that focus on modeling and/or measurement of air-to-ground wireless channels \cite{J_Wahab_Survey_2019, J_AlHourani_Optimal_2014, J_Matolak_Measure1_2017, J_Matolak_Measure2_2017, J_Matolak_Measure3_2017, J_Sabzehali_3D_2021, J_Cai_Empirical_2019, J_Liu_High_2021, C_Khawaja_UWB_2016, 3gpp_36777}. Air-to-air and air-to-ground channels are inherently different from ground-to-ground channels due to many reasons, such as the higher probability of LoS in the aerial channels and the mobility of UAVs \cite{J_Mozaffari_Tutorial_2018, J_Zeng_Accessing_2019, J_Wahab_Survey_2019}. Following the mathematical model suggested by the international telecommunication union (ITU) \cite{ITU}, perhaps the first work that established a meaningful yet simple relation between the LoS probability and the elevation angle in an aerial-terrestrial setup for low altitude platforms (LAPs) was \cite{J_AlHourani_Optimal_2014}, where the authors fitted a modified Sigmoid function to the LoS probability. Extensive measurement campaigns were conducted in \cite{J_Matolak_Measure1_2017, J_Matolak_Measure2_2017, J_Matolak_Measure3_2017} to obtain specific air-to-ground channel models for different environments (over-water \cite{J_Matolak_Measure1_2017}, hilly and mountainous \cite{J_Matolak_Measure2_2017}, and suburban and near-urban \cite{J_Matolak_Measure3_2017}). In these papers, the authors characterized different metrics related to the air-to-ground channels, such as path loss, Ricean $K$ factor of the small-scale fading, and delay spread of the channel. Another relatively recent measurement campaign was conducted in \cite{J_Cai_Empirical_2019}, where the authors extracted channel impulse responses from the received data, using which they obtained key characteristics of the channel, such as path loss and shadow fading. In \cite{J_Liu_High_2021}, the authors proposed an air-to-ground channel model for UAV base stations flying at high altitudes and moving periodically in circular curves. Considering ultrawideband signals, the authors in \cite{C_Khawaja_UWB_2016} conducted measurements for air-to-ground wireless channels and developed stochastic path-loss and multi-path channel models for ultrawideband propagation channels. Last but not the least, the third generation partnership project (3GPP) has also provided a detailed channel model, including the LoS probability and path-loss and fading models, in different environments in its technical report on cellular support for UAVs \cite{3gpp_36777}.

\textit{UAV Wobbling.}
One unique feature of aerial wireless communications, which can be viewed as an impairment, is the fact that UAVs may experience wobbling, i.e., typically small and random fluctuations of the UAV platform due to various reasons, such as wind gusts and lack of fixed infrastructure \cite{J_Ahmed_Flight_2010}. Since this area of research is still at its nascent stage, there are only a handful of works that incorporate UAV wobbling \cite{J_Xu_Multiuser_2020, J_Wu_Energy_2020, J_Dabiri_Analytical_2020, J_Morteza_Impact_2020, J_Pourranjbar_Novel_2020, J_Yang_Impact_2022, C_Yang_Impact_2022}. For instance, the authors in \cite{J_Xu_Multiuser_2020} provided a stochastic model for UAV wobbling and then studied a resource allocation problem in UAV-assisted cellular networks. Considering directional antennas, the antenna gain mismatch problem due to random UAV wobbling for an air-to-air network was studied in \cite{J_Dabiri_Analytical_2020}, where the authors also provided closed-form statistical channel models for the air-to-air links. In \cite{J_Morteza_Impact_2020}, the authors studied the impact of random UAV wobbling on the air-to-ground wireless channel and obtained the coherence time of the channel under different stochastic wobbling models. Following \cite{J_Morteza_Impact_2020}, the authors in \cite{J_Yang_Impact_2022} studied the Doppler effect at millimeter-wave (mmWave) frequencies for an air-to-ground channel under random UAV wobbling. Apart from analytical results, measurement campaigns have also reported the impact of wobbling as an important source of error in establishing a strong connection in aerial wireless networks \cite{C_Gomez_Air_2021, C_Semkin_Lightweight_2021}. In particular, in \cite{C_Gomez_Air_2021}, the authors used a channel sounder to investigate an air-to-ground wireless link, from which they observed that the variations of the received power from a completely static UAV are much less than that of a hovering UAV. These studies highlight the importance of considering UAV wobbling in aerial wireless communications.

\textit{Hardware Impairments.}
The unfavorable impact of RF imperfections on various aspects of communication systems has been studied extensively during the past decades. Perhaps one of the most comprehensive books on this subject is \cite{B_Schenk_Imperfections_2008}, where the authors describe in detail three fundamental hardware impairments, i.e., phase noise, I/Q imbalance, and PA nonlinearity, along with a general error model that characterizes the effect of all three impairments on the received signal in a wireless channel. This general error model \cite[Ch. 7]{B_Schenk_Imperfections_2008} motivated many researchers to redefine and solve fundamental communication-theoretic problems under a more realistic impairments-aware channel model \cite{J_Emil_New_2013, J_Emil_Capacity_2013, C_Studer_MIMO_2010, J_Emil_Massive_2014, C_Gustavsson_Impact_2014, J_Taghizadeh_Hardware_2018_1, J_Taghizadeh_Hardware_2018_2, J_Radhakrishnan_Hardware_2021, J_Li_Residual_2020}. For instance, the authors in \cite{J_Emil_New_2013} quantified the aggregate effect of hardware impairments on two-hop relaying systems \cite{J_Morteza_3D_2021, C_Morteza_Fundamentals_2021, J_Sabzehali_Optimizing_2021} and obtained closed-form expressions for the outage probability in these networks. In \cite{J_Emil_Capacity_2013}, the authors analytically proved that the capacity of a multiple-input multiple-output (MIMO) channel will be limited when hardware impairments are assumed, which signifies the fundamental impact of these impairments on the wireless channel. The problem of MIMO transmission with residual RF impairments on the transmitter side was studied in \cite{C_Studer_MIMO_2010}, where the authors showed that such hardware impairments substantially degrade the performance of MIMO detection algorithms. In \cite{J_Emil_Massive_2014}, the authors incorporated hardware impairments in the analysis of massive MIMO systems, and demonstrated the existence of a bound on the capacity of each user equipment (UE). Another interesting work that studied the impact of hardware impairments on massive MIMO systems is \cite{C_Gustavsson_Impact_2014}, where the authors perform two types of simulations considering additive and multiplicative stochastic impairment models, as well as more accurate deterministic behavioral models. 

Although sparse, there are a few works that study the impact of hardware impairments in UAV communications \cite{J_Li_UAV_2020, C_Hou_Hardware_2020}. For example, the authors in \cite{J_Li_UAV_2020} studied a UAV-aided non-orthogonal multiple access (NOMA) relaying network, where both the UAV-relay and terrestrial UEs suffer from residual hardware impairments. Quite surprisingly, none of the recent works consider the impact of UAV wobbling along with other impairments. In this paper, we analyze the joint impact of hardware impairments and UAV wobbling on the wireless channel and provide a unified impairments-aware air-to-ground channel model. We summarize our contributions next.

\subsection{Contributions} \label{subsec:Contributions}
This paper provides a unified model for the air-to-ground wireless channel that suffers from both UAV wobbling and hardware impairments. In particular, we assume that a hovering UAV (transmitter) communicates with a UE on the ground (receiver) in a multi-path environment with Rician fading. Both the UAV and the UE suffer from various hardware impairments, such as phase noise, I/Q imbalance, and PA nonlinearity. Apart from these hardware impairments, we assume that the UAV also suffers from random wobbling, i.e., unpredictable physical fluctuations due to not having a fixed aerial infrastructure. We model UAV wobbling using two different random processes, i.e., the Wiener process and the sinusoidal process, and characterize the properties of each process separately. As for the hardware impairments, we model the aggregate effect of all of them as two multiplicative and additive distortion processes and study them in both nonstationary and stationary scenarios. Although modeling these hardware impairments is known and well-accepted in the literature, the interplay between hardware impairments and UAV wobbling has not been studied yet. For this setup, our contributions are highlighted next.

\subsubsection{Unified Impairments-Aware Wireless Channel Model} \label{subsubsec:1}
We present a unified channel model in a multi-path Rician fading environment that accounts for both UAV wobbling and hardware impairments. Our proposed model consists of a modified channel impulse response and a distortion-plus-noise process, where the former is affected by UAV wobbling and only the multiplicative distortion part of the hardware impairments, while the latter is affected by UAV wobbling and both the multiplicative and additive distortion parts of the hardware impairments.

\subsubsection{Derivation and Evaluation of Key Channel Metrics} \label{subsubsec:2}
We rigorously derive channel ACFs for both nonstationary and stationary impairments. Using these results, we extensively study four key metrics that are used to characterize a wireless channel, i.e., (i) PDP, (ii) coherence time, (iii) coherence bandwidth, and (iv) PSD of the distortion-plus-noise process.

\subsubsection{Design Insights and Implications on Higher Frequencies} \label{subsubsec:3}
We demonstrate that even with stationary hardware impairments, the received signal could be nonstationary due to the oscillatory nature of UAV wobbling, a phenomenon that could only be observed in air-to-ground wireless communications. The analysis presented in this paper suggests a high sensitivity of key channel metrics to both UAV wobbling and hardware impairments, especially at higher carrier frequencies. A particularly noteworthy implication of our results is the degradation of channel coherence time at higher carrier frequencies because of these effects. To the best of our knowledge, this is the first work that proposes an impairments-aware unified channel model for UAV communications and characterizes the key metrics associated with the channel.

\begin{figure}[!t]
	\begin{minipage}[!t]{0.49\textwidth}
		\centering
		\captionof{table}{Acronyms used in this paper.}		\includegraphics[width=\columnwidth]{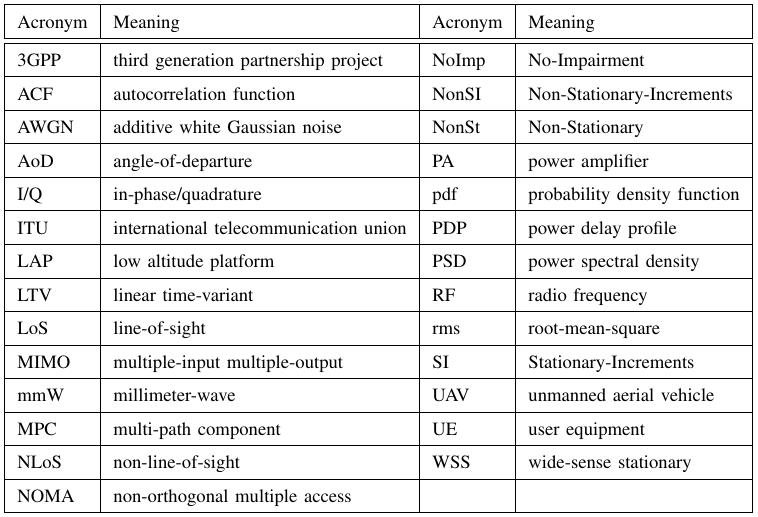}
		\label{tab:Table1}
	\end{minipage}
\vspace{-0.2cm}
	\hfill
	\begin{minipage}[!t]{0.49\textwidth}
		\centering
		\includegraphics[width=1\columnwidth]{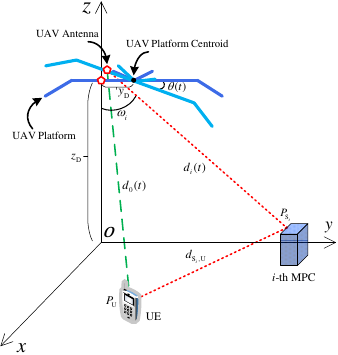}
		\captionof{figure}{Spatial setup: The UAV platform wobbles and the transceiver suffers from various hardware impairments.} 
		\label{fig:SystemModel}
	\end{minipage}
\end{figure}

For the reader's reference, list of the acronyms used in this paper is provided in Table \ref{tab:Table1}.

\section{System Model} \label{sec:SysMod}
\subsection{Spatial Setup} \label{subsec:SpatialSetup}
We consider a rotary-winged UAV that is hovering in the air and communicating with a ground UE. We assume that there are $N$ multi-path components (MPCs) from scatterers which contribute to the total received signal at the ground UE. To represent our system model, we use Cartesian coordinate system, where the ground is aligned with the $xy$-plane and the UAV platform centroid, ground UE, and the $i$-th scatterer (i.e., the $i$-th MPC) are located at $P_{\rm D} = (0, y_{\rm D}, z_{\rm D})$, $P_{\rm U} = (x_{\rm U}, y_{\rm U}, 0)$, and $P_{{\rm S}_i} = (x_{{\rm S}_i} , y_{{\rm S}_i} , z_{{\rm S}_i})$, respectively. As seen from Fig. \ref{fig:SystemModel}, we assume that the UAV is equipped with an antenna that is located at a distance of $y_{\rm D}$ from the centroid of the UAV platform, so that the initial location of the antenna is $(0, 0, z_{\rm D})$. Furthermore, for the ease of notation, we use the convention that the UE is the zeroth MPC. The UAV-UE, UAV-MPC, and MPC-UE distances are represented by $d_0(t)$, $d_i(t)$, and $d_{{\rm S}_i, {\rm U}}$ for the $i$-th MPC at time $t$, respectively. Also, the angle-of-departure (AoD) from the UAV to the $i$-th MPC (measured with respect to the $z$-axis) is denoted by $\omega_i$. Based on the defined convention, the AoD from the UAV to the UE is denoted by $\omega_0$.


\subsection{Impairments} \label{subsec:Impairments}
We categorize the nonidealities that degrade reliable aerial-terrestrial communications as \textit{UAV wobbling} and \textit{hardware impairments}, which are introduced next. 

\textit{UAV Wobbling}: The agility of UAVs comes at the price of not having a fixed infrastructure, which could even lead to instability because of wind gusts or bad weather conditions \cite{J_Dabiri_Analytical_2020}. Considering these natural phenomena and the high vibration frequency of UAVs' propellers, UAVs may experience small random physical vibrations, which are referred to as wobbling in this paper. Although this wobbling is typically small \cite{J_Ahmed_Flight_2010}, its impact on the received signals could be non-negligible especially at high carrier frequencies. Furthermore, this impact could be even more severe if the UAV is equipped with a directional antenna instead of an omnidirectional antenna. As shown in Fig. \ref{fig:SystemModel}, we represent the wobbling pitch angle at time $t$ by $\theta(t)$, which shows the deviation of the UAV platform from its initial state. In this paper, we model $\theta(t)$ using stochastic processes \cite{J_Morteza_Impact_2020} and study the impact of UAV wobbling on different aspects of the wireless channel. More technical discussions are presented in Section \ref{subsec:Eval}.

\textit{Hardware Impairments}: Because of their inherent physical characteristics, RF transceivers are known to suffer from hardware impairments, such as PA nonlinearity, I/Q imbalance, and phase noise \cite{B_Schenk_Imperfections_2008, C_Studer_MIMO_2010}. Although various models for each hardware impairment can be found in the literature \cite{C_Schenk_Impact_2006, J_Schenk_Performance_2007, J_Piazzo_Analysis_2002}, we intend to consider the aggregate effect of all such impairments in this work for the sake of generality. The fundamental impact of each of these impairments on the transmitted and received signals usually includes an attenuation and phase rotation of the original signal (i.e., multiplicative effect) along with a leakage of different carriers or symbols into each other (i.e., additive effect). Due to these factors, the influence of each hardware impairment can be modeled as an affine linear function. Therefore, the combined effect of these hardware impairments can be modeled as two multiplicative and additive distortion noise processes at both the transmitter and receiver \cite[Sec. 7.2.2]{B_Schenk_Imperfections_2008}. Mathematically speaking, the \textit{impaired} transmitted and received signals in complex baseband can be written as
\begin{align}
s(t) = \chi_{\rm T}(t) \tilde{s}(t) + \eta_{\rm T}(t), \quad r(t) = \chi_{\rm R}(t) \tilde{r}(t) + \eta_{\rm R}(t), \label{s_HI} 
\end{align}
where $\tilde{s}(t)$ and $\tilde{r}(t)$ are the \textit{unimpaired} transmitted and received signals, $s(t)$ and $r(t)$ are their impaired counterparts, and $\chi_{\rm T}(t)$ and $\eta_{\rm T}(t)$ (resp. $\chi_{\rm R}(t)$ and $\eta_{\rm R}(t)$) are the multiplicative and additive distortion noise processes at the transmitter (resp. receiver), respectively. Note that these RF impairments are nonstationary in nature, and thus, these distortion noises are assumed to be time-dependent in general \cite[Sec. 7.2.3]{B_Schenk_Imperfections_2008}. These four noise processes are characterized by their ACFs, i.e., $A_{\chi_{\rm T}}(t_1, t_2) = \nbbE\left[ \chi_{\rm T}^*(t_1) \chi_{\rm T}(t_2) \right]$, $A_{\chi_{\rm R}}(t_1, t_2) = \nbbE\left[ \chi_{\rm R}^*(t_1) \chi_{\rm R}(t_2) \right]$, $A_{\eta_{\rm T}}(t_1, t_2) = \nbbE\left[ \eta_{\rm T}^*(t_1) \eta_{\rm T}(t_2) \right]$, and $A_{\eta_{\rm R}}(t_1, t_2) = \nbbE\left[ \eta_{\rm R}^*(t_1) \eta_{\rm R}(t_2) \right]$. Furthermore, we assume that these four processes are independent from each other and the additive distortions have zero means \cite{B_Schenk_Imperfections_2008, J_Emil_New_2013}, i.e., $\nbbE\left[ \eta_{\rm T}(t) \right] = \nbbE\left[ \eta_{\rm R}(t) \right] = 0$. Since WSS processes are an important class of random processes, we will also study our channel metrics (described next) when hardware impairments are assumed to be WSS. In that case, the above-mentioned ACFs will only be functions of $t_2 - t_1$.


\subsection{Preliminaries and Assumptions} \label{subsec:Prelim}
Before presenting our channel model, we provide some preliminary results and review the well-established assumptions and properties of the unimpaired channel model \cite[Ch. 3]{B_Goldsmith_Wireless_2005}. We start by establishing a connection between UAV-MPC distances at different times and the wobbling pitch angle, presented next (see \cite[Lemma 1]{J_Morteza_Impact_2020} for the proof).
\begin{lemma} \label{lem:DiffDist}
	Assuming that $y_{\rm D} \ll d_i(t)$ and $\theta(t) \ll 1~{\rm rad}$ for a wobbling UAV, we have
	\begin{align}\label{dDn}
		d_i(t+\Delta t) - d_i(t) \approx y_{\rm D}\cos(\omega_i)[\theta(t+\Delta t) - \theta(t)].
	\end{align}
\end{lemma}
\begin{cor}\label{cor1}
	If $\theta(t)$ is a random process with stationary increments, i.e., $\theta(t + \Delta t) - \theta(t)$ and $\theta(\Delta t)$ have the same distribution, we have $d_i(t+\Delta t) - d_i(t) \approx y_{\rm D}\cos(\omega_i)\theta(\Delta t)$.
\end{cor}
Since the distance between the transmitter and receiver changes with time due to UAV wobbling, we observe a Doppler shift in the channel. Specifically, the Doppler phase shift for the $i$-th MPC can be written using \eqref{dDn} as 
\begin{align}\label{Doppler}
	\varphi_{{\rm D}_i}(t, \Delta t) = \frac{2\pi}{\lambda} y_{\rm D}\cos(\omega_i)[\theta(t+\Delta t) - \theta(t)],
\end{align}
where $\lambda = \frac{c}{f_{\rm c}}$ is the signal wavelength, $c$ is the speed of light, and $f_{\rm c}$ is the carrier frequency.

We consider a multi-path channel with one LoS and $N$ non-line-of-sight (NLoS) links. In Fig. \ref{fig:SystemModel}, the LoS path from the UAV to the ground UE is represented by a green dashed line and the NLoS path from the UAV to the $i$-th MPC and then to the ground UE is represented by red dotted lines. Without hardware impairments, the received signal in the complex baseband and the channel impulse response can be written, respectively, as \cite{B_Goldsmith_Wireless_2005}
\begin{align}
	\tilde{r}(t) &= \sum_{i=0}^{N} \alpha_i(t) {\rm e}^{-{\rm j}\varphi_i(t)} \tilde{s}(t - \tau_i(t)) + \tilde{n}(t),\notag\\
	\tilde{c}(\tau; t) &= \sum_{i=0}^{N} \alpha_i(t) {\rm e}^{-{\rm j}\varphi_i(t)} \delta(\tau - \tau_i(t)) \label{c_noHI},
\end{align}
where $\alpha_i(t)$, $\varphi_i(t)$, and $\tau_i(t)$ are the amplitude, phase, and delay of the $i$-th MPC, respectively, $\tilde{n}(t)$ is the AWGN, $\tilde{c}(\tau; t)$ is the unimpaired channel impulse response, and $\delta(.)$ is the Dirac delta function. From \eqref{c_noHI}, one can easily verify the following convolution integral for the channel, which is considered as a linear time-variant (LTV) system:
\begin{align} \label{LTV_noHI}
	\tilde{r}(t) = \int_{-\infty}^{\infty} \tilde{c}(\tau; t) \tilde{s}(t-\tau)\,{\rm d}\tau + \tilde{n}(t).
\end{align}

We assume that $\alpha_i(t)$ does not change considerably over the time period of interest, and thus, drop its $t$ argument henceforth. We also consider the well-accepted Laplacian model for $|\alpha_i|^2$ as
\begin{align} \label{Laplace}
	|\alpha_i|^2 = \frac{1}{2\beta} {\rm e}^{-\frac{|\omega_i - \omega_0|}{\beta}}, \qquad 1 \leq i \leq N,
\end{align}
where $\beta > 0$ is a scaling parameter \cite{J_Pedersen_Power_1997, B_Molisch_Wireless_2011} and for the AoD we assume that $\omega_i \sim U[0, \frac{\pi}{2})$. Therefore, the power of the $i$-th MPC is derived as $P_{\alpha_i} \coloneqq \nbbE\left[|\alpha_i|^2\right] = \frac{1}{2\pi}\left( 2 - {\rm e}^{-\frac{\omega_0}{\beta}} - {\rm e}^{-\frac{\pi/2 - \omega_0}{\beta}} \right)$. Furthermore, assuming Rician fading model with factor $K$, we can write $|\alpha_0|^2 = K \sum_{i=1}^N |\alpha_i|^2 = K \sum_{i=1}^N \frac{1}{2\beta}{\rm e}^{-\frac{|\omega_i - \omega_0|}{\beta}}$, which gives the power of the LoS component as $P_{\alpha_0} \coloneqq \nbbE\left[|\alpha_0|^2\right] = N K P_{\alpha_i}$.

\begin{remark}
Instead of using the Laplacian model, we could have used more elaborate 3GPP-based path loss models \cite[Table B-2]{3gpp_36777}. However, since our fundamental results (provided next) are agnostic to the choice of the channel model, we favored the current channel model assumptions that are equally well accepted in the literature \cite{B_Goldsmith_Wireless_2005, J_Pedersen_Power_1997, B_Molisch_Wireless_2011} and also mathematically more tractable than the 3GPP-based models.
\end{remark}

The delay of the $i$-th MPC can be written as $\tau_i(t) \approx \tau_i(0) = \frac{d_i(0) + d_{{\rm S}_i, {\rm U}}}{c}$, where we assumed that the delay also does not change significantly over time. In fact, since the UAV-UE distance is in the order of hundreds of meters or even kilometers, the delay will be in the order of microseconds. However, using \eqref{dDn}, it is clear that the added distance due to UAV wobbling is in the order of centimeters, which makes the corresponding residual delay to be in the order of nanoseconds. Hence, our approximation is valid and we also define $\tau_i \coloneqq \tau_i(0)$ for simplicity. Assuming that the location of the UE ($P_{\rm U}$) is known and the MPCs are distributed randomly on the ground, the propagation delay of the $i$-th MPC can be written as $\tau_i = \tau_0 + \tau_{\Delta i}$, where $\tau_0 = \frac{d_0}{{\rm c}}$ is the UAV-UE delay and $\tau_{\Delta i}$ is the excess delay of the $i$-th MPC. We assume that $\tau_{\Delta i}$ is distributed exponentially with parameter $\rho_i$, which depends on the propagation environment \cite{B_Molisch_Wireless_2011, J_Lone_Characterizing_2022}. Therefore, the probability density functions (pdfs) of $\tau_0$ and $\tau_i$ can be written as
\begin{align} \label{f_tau}
	f_{\tau_0}(\tau) = \delta(\tau - \tau_0), \quad f_{\tau_i}(\tau) = \rho_i{\rm e}^{-\rho_i(\tau - \tau_0)} {\bf 1}(\tau - \tau_0),
\end{align}
where ${\bf 1}(.)$ is the indicator function.

As for the phase of the $i$-th MPC, on the other hand, the small distance variations described earlier become more important since they will be multiplied by the carrier frequency, causing Doppler phase shift. We write the phase of the $i$-th MPC as
\begin{align} \label{phi_i}
	\varphi_i(t) &= \frac{2\pi}{\lambda} \left(d_i(0) + d_{{\rm S}_i, {\rm U}}\right) + \frac{2\pi}{\lambda} y_{\rm D}\cos(\omega_i)[\theta(t) - \theta(0)]\notag\\
	&= 2\pi f_{\rm c} \tau_i + \varphi_{{\rm D}_i}(t),
\end{align}
where $\varphi_{{\rm D}_i}(t) \coloneqq \varphi_{{\rm D}_i}(0, t)$ is the Doppler phase shift at time $t$ given in \eqref{Doppler}.

\subsection{Metrics} \label{subsec:Metrics}
We consider multiple key channel-related metrics in this paper, which are described next. We represent the channel impulse response by $c(\tau; t)$, where $\tau$ is the delay variable and $t$ is the observation time, and define its ACF as $A_{\rm c}(\tau; t, \Delta t) = \mathbb{E}\left[c^*(\tau; t)c(\tau; t + \Delta t)\right]$. Note that since the channel is assumed to be nonstationary, we present our definitions/results in their general form. Throughout the paper, we assume that the channel undergoes uncorrelated scattering.

\subsubsection{PDP} \label{subsubsec:PDP}
We define PDP in a nonstationary multi-path channel as the average received power at a given multi-path delay $\tau$ and a specific observation time $t$ \cite{B_Goldsmith_Wireless_2005, B_Molisch_Wireless_2011}. One way to obtain the PDP is to evaluate the ACF of the channel impulse response at $\Delta t = 0$, i.e., $P_{\rm c}(\tau; t) = A_{\rm c}(\tau; t, \Delta t = 0) = \mathbb{E}\left[|c(\tau; t)|^2\right]$. However, as we will see in the next section, the channel impulse response entails a sum of Dirac delta functions (see \eqref{c_noHI} and \eqref{c_HI}), which introduces a singularity problem (delta squared) into this definition. One way to avoid this issue is to isolate one of the delta functions by writing the channel ACF as $A_{\rm c}(\tau; t, \Delta t) = P_{\rm c}(\tau; t) \delta(\Delta t)$, and \textit{redefining} the PDP with only one delta function as $P_{\rm c}(\tau; t)$ \cite{J_Meijerink_Physical_2014}.

\subsubsection{Coherence Time} \label{subsubsec:CT}
Coherence time is defined as the period of time over which the channel remains almost constant, calculated as follows: First, we evaluate the ACF of the channel impulse response, i.e., $A_{\rm c}(\tau; t, \Delta t)$. Note that for a nonstationary channel, this ACF is a function of $\tau$, $t$, and $\Delta t$. Then, we take the Fourier transform of this ACF with respect to the delay variable $\tau$ and represent the new frequency variable by $\Delta f$, i.e., $A_{\rm C}(\Delta f; t, \Delta t) = \ncalF_\tau\{A_{\rm c}(\tau; t, \Delta t)\}$. Now, setting $\Delta f = 0$ and normalizing this ACF, we define coherence time as
\begin{align} \label{CT}
	T_{\rm Coh}(t) \coloneqq \min\left\{\Delta t: \frac{|A_{\rm C}(t, \Delta t)|}{\max{|A_{\rm C}(t, \Delta t)|}} \leq \gamma_{\rm T}\right\},
\end{align}
where $A_{\rm C}(t, \Delta t) = A_{\rm C}(\Delta f = 0; t, \Delta t)$ and $\gamma_{\rm T}$ is a predefined threshold \cite{C_Va_Basic_2015, J_Morteza_Impact_2020}.

\subsubsection{Coherence Bandwidth} \label{subsubsec:CB}
Coherence bandwidth is defined as the range of frequencies over which the channel remains almost constant, calculated as follows: We first obtain the Fourier transform of the channel ACF as we did in the definition of coherence time, i.e., $A_{\rm C}(\Delta f; t, \Delta t)$. Then, setting $\Delta t = 0$ and normalizing this ACF, we define coherence bandwidth as
\begin{align} \label{CB}
	B_{\rm Coh}(t) \coloneqq \min\left\{\Delta f: \frac{|A_{\rm C}(\Delta f; t)|}{\max{|A_{\rm C}(\Delta f; t)|}} \leq \gamma_{\rm B}\right\},
\end{align}
where $A_{\rm C}(\Delta f; t) = A_{\rm C}(\Delta f; t, \Delta t = 0)$ and $\gamma_{\rm B}$ is a predefined threshold.


\subsubsection{PSD of Distortion-Plus-Noise} \label{subsubsec:PSD}
As seen already, UAV wobbling and hardware impairments will cause both multiplicative and additive distortions to the received signal. In this paper, we treat the additive distortion as noise, and together with additive white Gaussian noise (AWGN), study their PSDs as our last metric of interest. The PSD of nonstationary signal $x(t)$ is defined as the Fourier transform of its time-averaged ACF, i.e., $S_{x}(f) = \ncalF_{\Delta t}\{\langle\nbbE[x^*(t)x(t + \Delta t)]\rangle_t\}$,
where $\langle y(t)\rangle_t = \lim_{T\to \infty}\frac{1}{2T}\int_{-T}^{T}y(t)\,{\rm d}t$ represents averaging $y(t)$ over time \cite[Sec. 7.2]{B_Peebels_Probability_2001}.

\section{Impairments-Aware Unified Channel Model} \label{sec:Unified}

\subsection{Unified Channel Model} \label{subsec:Unified}
Now, let us revisit \eqref{c_noHI} when there exist hardware impairments. Using \eqref{s_HI}, we can write the received signal in complex baseband as
\begin{align} \label{r_HI}
r(t) &= \chi_{\rm R}(t) \left[\sum_{i=0}^{N} \alpha_i(t) {\rm e}^{-{\rm j}\varphi_i(t)} s(t - \tau_i(t))\right] + \eta_{\rm R}(t) + \tilde{n}(t) \vspace{-0.0cm} \notag \\
&= \chi_{\rm R}(t) \Bigg[\sum_{i=0}^{N} \alpha_i(t) {\rm e}^{-{\rm j}\varphi_i(t)} \Big( \chi_{\rm T}(t - \tau_i(t)) \tilde{s}(t - \tau_i(t))\notag\\
&\quad + \eta_{\rm T}(t - \tau_i(t)) \Big)\Bigg] + \eta_{\rm R}(t) + \tilde{n}(t) \vspace{-0.0cm} \notag \\
&= \sum_{i=0}^{N} \!\alpha_i(t) {\rm e}^{-{\rm j}\varphi_i(t)} \chi_{\rm R}(t) \chi_{\rm T}(t - \tau_i(t))\tilde{s}(t - \tau_i(t)) \!+\! n(t),
\end{align}
where $n(t)$ is the combined effect of AWGN, wobbling, and hardware impairments, given as
\begin{align} \label{n_HI}
n(t) = \sum_{i=0}^{N} \alpha_i(t) {\rm e}^{-{\rm j}\varphi_i(t)} \chi_{\rm R}(t) \eta_{\rm T}(t - \tau_i(t)) + \eta_{\rm R}(t) + \tilde{n}(t).
\end{align}
From \eqref{r_HI}, we obtain the channel impulse response as
\begin{align} \label{c_HI}
c(\tau; t) = \sum_{i=0}^{N} \alpha_i(t) {\rm e}^{-{\rm j}\varphi_i(t)} \chi_{\rm R}(t) \chi_{\rm T}(t - \tau_i(t))\delta(\tau - \tau_i(t)),
\end{align}
where the impact of UAV wobbling is hidden in the phase term, i.e., $\varphi_i(t)$ (see \eqref{phi_i}), while hardware impairments manifest their effect as multiplicative distortion noises, i.e., $\chi_{\rm R}(t)$ and  $\chi_{\rm T}(t - \tau_i(t))$. This modified channel impulse response together with the new noise term in \eqref{n_HI}, also referred to as the distortion-plus-noise process in this paper, construct our impairments-aware unified channel model for air-to-ground wireless communications. Completely analogous to the convolution integral for the without-impairment scenario in \eqref{LTV_noHI}, we can write the input-output relation of the new channel (which is also an LTV system) as $r(t) = \int_{-\infty}^{\infty} c(\tau; t) \tilde{s}(t-\tau)\,{\rm d}\tau + n(t)$.
Note that the input to the channel will still be the unimpaired signal $\tilde{s}(t)$.
\subsection{Metrics: General Results} \label{subsec:General}
In order to further characterize this unified channel model, we obtain the metrics defined in Section \ref{subsec:Metrics} next. The following lemma provides the channel ACF in the time-delay domain.
\begin{lemma} \label{lem:ACF}
For a nonstationary air-to-ground wireless channel with physical nonidealities (due to UAV wobbling) and hardware impairments (due to intrinsic RF components), where the channel impulse response is given by \eqref{c_HI}, the channel ACF can be written as
\begin{align}\label{ACF_nonWSS}
&A_{\rm c}^{\rm NonSt}(\tau; t, \Delta t) = \sum_{i=0}^{N} A_{\chi_{\rm R}}(t, t \!+\! \Delta t) A_{\chi_{\rm T}}(t \!-\! \tau, t \!-\! \tau \!+\! \Delta t) \notag \\
&\times\!\nbbE\!\left[ |\alpha_i|^2\! \exp\left\{-{\rm j} \frac{2\pi}{\lambda} y_{\rm D}\cos(\omega_i)\!\left[\theta(t \!+\! \Delta t) \!- \!\theta(t)\right]\right\} \right]\! f_{\tau_i}(\tau),
\end{align}
where the superscript ${\rm NonSt}$ stands for ``Non-Stationary" and $f_{\tau_i}(\tau)$ represents the pdf of $\tau_i$.
\end{lemma}
\begin{IEEEproof}
We write the channel ACF as
\begin{align*}
&A_{\rm c}^{\rm NonSt}(\tau; t, \Delta t) = \mathbb{E}\left[c^*(\tau; t)c(\tau; t + \Delta t)\right] \\
&= \mathbb{E}\Bigg[ \sum_{i=0}^{N} \sum_{k=0}^{N} \alpha_i^*\alpha_k {\rm e}^{-{\rm j}(\varphi_k(t + \Delta t) - \varphi_i(t))} \chi_{\rm R}^*(t)\chi_{\rm R}(t + \Delta t) \vspace{-0.0cm} \\
&\qquad\times\chi_{\rm T}^*(t - \tau_i)\chi_{\rm T}(t - \tau_k + \Delta t)\delta(\tau - \tau_i)\delta(\tau - \tau_k) \Bigg]\vspace{-0.0cm}\\
&= \sum_{i=0}^{N} \sum_{\substack{k=0\\k\neq i}}^{N} \mathbb{E}\Big[ \alpha_i^*\alpha_k {\rm e}^{-{\rm j}\frac{2\pi}{\lambda} \left(d_k(0) + d_{{\rm S}_k, {\rm U}} - d_i(0) - d_{{\rm S}_i, {\rm U}}\right)}\vspace{-0.0cm} \\
&\qquad\times {\rm e}^{-{\rm j}\left(\varphi_{{\rm D}_k}(t + \Delta t) - \varphi_{{\rm D}_i}(t)\right)}\chi_{\rm R}^*(t)\chi_{\rm R}(t + \Delta t) \\
&\qquad\times \chi_{\rm T}^*(t - \tau_i)\chi_{\rm T}(t - \tau_k + \Delta t)\delta(\tau - \tau_i)\delta(\tau - \tau_k) \Big]\vspace{-0.0cm}\\
&\quad+ \sum_{i=0}^{N} \mathbb{E}\Big[ |\alpha_i|^2 {\rm e}^{-{\rm j}\left(\varphi_{{\rm D}_i}(t + \Delta t) - \varphi_{{\rm D}_i}(t)\right)} \chi_{\rm R}^*(t)\chi_{\rm R}(t \!+\! \Delta t)\\
&\qquad\times \chi_{\rm T}^*(t \!-\! \tau_i)\chi_{\rm T}(t \!-\! \tau_i \!+ \!\Delta t)\delta(\tau \!-\! \tau_i)\delta(\tau \!-\! \tau_i) \Big].\vspace{-0.0cm}
\end{align*}
Since $d_{{\rm S}_k, {\rm U}} - d_{{\rm S}_i, {\rm U}} \gg \lambda$, we observe that $Z_{k,i}=\left[\frac{2\pi}{\lambda} \left(d_{{\rm S}_k, {\rm U}} - d_{{\rm S}_i, {\rm U}}\right) {\rm mod~} 2\pi\right]$ is a uniformly distributed random variable from $0$ to $2\pi$ \cite[Lemma 4]{J_Dhillon_Backhaul_2015}. Therefore, $\nbbE\left[{\rm e}^{-{\rm j}Z_{k,i}}\right] = 0$ because of which the double-summation in the last equality is zero. As for the single summation term, we encounter the singularity issue (delta squared) discussed earlier in Sec. \ref{subsubsec:PDP}. To avoid this singularity problem, we isolate one of the delta functions \cite{J_Meijerink_Physical_2014} and rewrite the channel ACF with only one delta function as $A_{\rm c}^{\rm NonSt}(\tau; t, \Delta t)$
\begin{align*}
&= \sum_{i=0}^{N} \mathbb{E}\Big[ |\alpha_i|^2 {\rm e}^{-{\rm j}\frac{2\pi}{\lambda} y_{\rm D}\cos(\omega_i)\left[\theta(t + \Delta t) - \theta(t)\right]}\\
&\quad\times\chi_{\rm R}^*(t)\chi_{\rm R}(t \!+\! \Delta t) \chi_{\rm T}^*(t \!-\! \tau_i)\chi_{\rm T}(t \!-\! \tau_i \!+\! \Delta t)\delta(\tau \!-\! \tau_i)\Big]\\
&= \sum_{i=0}^{N} \int_{-\infty}^{\infty} \nbbE\left[|\alpha_i|^2 {\rm e}^{-{\rm j}\frac{2\pi}{\lambda} y_{\rm D}\cos(\omega_i)\left[\theta(t + \Delta t) - \theta(t)\right]}\right] \\
&\quad\times A_{\chi_{\rm R}}(t, t \!+\! \Delta t)  A_{\chi_{\rm T}}(t \!-\! \tau_i, t \!-\! \tau_i \!+\! \Delta t) \delta(\tau \!-\! \tau_i)f_{\tau_i}(\tau_i){\rm d}\tau_i,
\end{align*}
which gives the result in \eqref{ACF_nonWSS} by applying the sifting property of the delta function.
\end{IEEEproof}
The following corollary simplifies \eqref{ACF_nonWSS} further by considering WSS hardware impairments.
\begin{cor} \label{cor2}
Assuming WSS hardware impairments, the channel ACF can be simplified as
\begin{align}\label{ACF_WSS}
&A_{\rm c}^{\rm WSS}(\tau; t, \Delta t) = \sum_{i=0}^{N} A_{\chi_{\rm R}}(\Delta t) A_{\chi_{\rm T}}(\Delta t)\notag\\
&\quad\times \nbbE\left[ |\alpha_i|^2 {\rm e}^{-{\rm j} \frac{2\pi}{\lambda} y_{\rm D}\cos(\omega_i)\left[\theta(t + \Delta t) - \theta(t)\right]} \right] f_{\tau_i}(\tau).
\end{align}
Note that the superscript ${\rm WSS}$ only pertains to ``hardware" impairments, not UAV wobbling. In addition, assuming that $\theta(t)$ is a random process with stationary increments, we have 
\begin{align}\label{ACF_WSS_StInc}
&A_{\rm c}^{\rm WSS-SI}(\tau; \Delta t) = \sum_{i=0}^{N} A_{\chi_{\rm R}}(\Delta t) A_{\chi_{\rm T}}(\Delta t)\notag\\
&\quad\times \nbbE\left[ |\alpha_i|^2 {\rm e}^{-{\rm j} \frac{2\pi}{\lambda} y_{\rm D}\cos(\omega_i)\theta(\Delta t)} \right] f_{\tau_i}(\tau),
\end{align}
which will not be a function of $t$ anymore. Thus, the channel in this special case becomes stationary. The abbreviation ${\rm SI}$ in the superscript stands for ``Stationary-Increments".
\end{cor}
The following theorem gives the channel PDP for both nonstationary and WSS impairments.
\begin{theorem}\label{thm1}
The PDP of the air-to-ground wireless channel for the nonstationary and WSS hardware impairments can be formulated, respectively, as
\begin{align}
P_{\rm c}^{\rm NonSt}(\tau; t) &= \sum_{i=0}^{N} P_{\alpha_i} P_{\chi_{\rm R}(t)} P_{\chi_{\rm T}(t - \tau)} f_{\tau_i}(\tau),\notag\\
P_{\rm c}^{\rm WSS}(\tau) &= \sum_{i=0}^{N} P_{\alpha_i} P_{\chi_{\rm R}} P_{\chi_{\rm T}} f_{\tau_i}(\tau), \label{PDP_WSS}
\end{align}
where $P_{\chi_{\rm R}(t)} \coloneqq A_{\chi_{\rm R}}(t, t) = \nbbE\left[|\chi_{\rm R}(t)|^2\right]$ and $P_{\chi_{\rm T}(t - \tau)} \coloneqq A_{\chi_{\rm T}}(t - \tau, t - \tau) = \nbbE\left[|\chi_{\rm T}(t - \tau)|^2\right]$ are the powers of the multiplicative distortion noises at the receiver and transmitter at times $t$ and $t - \tau$, respectively. The definitions of their WSS counterparts are similar, i.e., $P_{\chi_{\rm R}} \coloneqq \nbbE\left[|\chi_{\rm R}|^2\right]$ and $P_{\chi_{\rm T}} \coloneqq \nbbE\left[|\chi_{\rm T}|^2\right]$. Clearly, we also have $P_{\rm c}^{\rm WSS-SI}(\tau) = P_{\rm c}^{\rm WSS}(\tau)$.
\end{theorem}
\begin{IEEEproof}
Setting $\Delta t = 0$ in \eqref{ACF_nonWSS} and \eqref{ACF_WSS}, we get \eqref{PDP_WSS}.
\end{IEEEproof}
Note that UAV wobbling does not play any role in determining the PDP. Further, from \eqref{PDP_WSS}, we observe that the PDP will be proportional to the pdfs of the propagation delays of different MPCs \cite[p. 348]{B_Patzold_Mobile_2012}. Using Theorem \ref{thm1}, we can also determine the average and root-mean-square (rms) delay spreads for nonstationary and WSS hardware impairments, respectively, as
\begin{align}
\mu^{\rm NonSt}(t) &= \frac{\int_{0}^{\infty} \tau P_{\rm c}(\tau; t) \, {\rm d}\tau}{\int_{0}^{\infty} P_{\rm c}(\tau; t) \, {\rm d}\tau} \notag\\
&= \frac{\sum_{i=0}^{N} P_{\alpha_i} \int_{0}^{\infty} \tau P_{\chi_{\rm T}(t - \tau)} f_{\tau_i}(\tau)  \, {\rm d}\tau}{\sum_{i=0}^{N} P_{\alpha_i} \int_{0}^{\infty} P_{\chi_{\rm T}(t - \tau)} f_{\tau_i}(\tau) \, {\rm d}\tau}, \label{mu-noWSS} \\
\sigma^{\rm NonSt}(t) &= \sqrt{\frac{\int_{0}^{\infty} (\tau - \mu(t))^2 P_{\rm c}(\tau; t) \, {\rm d}\tau}{\int_{0}^{\infty} P_{\rm c}(\tau; t) \, {\rm d}\tau}} \notag\\
&= \sqrt{\frac{\sum_{i=0}^{N} \!P_{\alpha_i} \!\int_{0}^{\infty} (\tau \!-\! \mu(t))^2 P_{\chi_{\rm T}(t - \tau)} f_{\tau_i}(\tau) \, {\rm d}\tau}{\sum_{i=0}^{N} P_{\alpha_i} \int_{0}^{\infty} P_{\chi_{\rm T}(t - \tau)} f_{\tau_i}(\tau) {\rm d}\tau}}, \label{sigma-noWSS} \\
\mu^{\rm WSS} &= \frac{\int_{0}^{\infty} \tau P_{\rm c}^{\rm WSS}(\tau) \, {\rm d}\tau}{\int_{0}^{\infty} P_{\rm c}^{\rm WSS}(\tau) \, {\rm d}\tau} \notag\\
&= \frac{\sum_{i=0}^{N} P_{\alpha_i} \int_{0}^{\infty} \tau f_{\tau_i}(\tau)  \, {\rm d}\tau}{\sum_{i=0}^{N} P_{\alpha_i}}, \label{mu-WSS} \\
\sigma^{\rm WSS} &= \sqrt{\frac{\int_{0}^{\infty} (\tau - \mu^{\rm WSS})^2 P_{\rm c}^{\rm WSS}(\tau) \, {\rm d}\tau}{\int_{0}^{\infty} P_{\rm c}^{\rm WSS}(\tau) \, {\rm d}\tau}} \notag\\
&= \sqrt{\frac{\sum_{i=0}^{N} P_{\alpha_i} \int_{0}^{\infty} (\tau - \mu^{\rm WSS})^2 f_{\tau_i}(\tau) \, {\rm d}\tau}{\sum_{i=0}^{N} P_{\alpha_i}}}. \label{sigma-WSS}
\end{align}

As for the coherence time/bandwidth, we need to first derive the Fourier transform of the channel ACF with respect to the delay variable $\tau$. For nonstationary impairments, we have
\begin{align}\label{F_ACF_nonWSS}
&A_{\rm C}^{\rm NonSt}(\Delta f; t, \Delta t) = \ncalF_{\tau}\{A_{\rm c}^{\rm NonSt}(\tau; t, \Delta t)\} \notag \\
&\quad =\sum_{i=0}^{N}\! A_{\chi_{\rm R}}(t, t \!+\! \Delta t)\nbbE\left[ |\alpha_i|^2 {\rm e}^{-{\rm j} \frac{2\pi}{\lambda} y_{\rm D}\cos(\omega_i)\left[\theta(t + \Delta t) - \theta(t)\right]} \right] \notag \\
&\qquad \times\!\int_{0}^{\infty}\! A_{\chi_{\rm T}}(t - \tau, t - \tau + \Delta t) f_{\tau_i}(\tau) {\rm e}^{-{\rm j}2\pi \Delta f \tau}\,{\rm d}\tau,
\end{align}
while for the WSS hardware impairments, \eqref{F_ACF_nonWSS} can be simplified as
\begin{align}\label{F_ACF_WSS}
&A_{\rm C}^{\rm WSS}(\Delta f; t, \Delta t)=\sum_{i=0}^{N} A_{\chi_{\rm R}}(\Delta t) A_{\chi_{\rm T}}(\Delta t)\notag\\
&\times\!\nbbE\left[ |\alpha_i|^2 {\rm e}^{-{\rm j} \frac{2\pi}{\lambda} y_{\rm D}\cos(\omega_i)\left[\theta(t + \Delta t) - \theta(t)\right]} \right]\!\! \int_{0}^{\infty}\!\!\! f_{\tau_i}(\tau) {\rm e}^{-{\rm j}2\pi \Delta f \tau}{\rm d}\tau.
\end{align}
Note that when $\theta(t)$ has the stationary-increments property, \eqref{F_ACF_WSS} is further simplified as
\begin{align}\label{F_ACF_WSS_SI}
&A_{\rm C}^{\rm WSS-SI}(\Delta f; \Delta t) = \sum_{i=0}^{N} A_{\chi_{\rm R}}(\Delta t) A_{\chi_{\rm T}}(\Delta t)\notag\\
&\times\! \nbbE\left[ |\alpha_i|^2 {\rm e}^{-{\rm j} \frac{2\pi}{\lambda} y_{\rm D}\cos(\omega_i)\theta(\Delta t)} \right] \!\int_{0}^{\infty}\!\! f_{\tau_i}(\tau) {\rm e}^{-{\rm j}2\pi \Delta f \tau}{\rm d}\tau,
\end{align}
which is not a function of $t$ anymore. The following two theorems give the coherence time and coherence bandwidth of the channel, respectively.



\begin{theorem} \label{thm2}
The coherence time of the air-to-ground channel can be formulated as in \eqref{CT}, where the ACFs for nonstationary and WSS hardware impairments are given, respectively, as
\begin{align}
A_{\rm C}^{\rm NonSt}(t, \Delta t) &=\sum_{i=0}^{N} A_{\chi_{\rm R}}(t, t + \Delta t)\notag\\
&\times\nbbE\left[ |\alpha_i|^2 {\rm e}^{-{\rm j} \frac{2\pi}{\lambda} y_{\rm D}\cos(\omega_i)\left[\theta(t + \Delta t) - \theta(t)\right]} \right] \notag\\
&\times\int_{0}^{\infty} \!\!A_{\chi_{\rm T}}(t \!-\! \tau, t \!- \!\tau \!+\! \Delta t) f_{\tau_i}(\tau)\,{\rm d}\tau, \label{AC_t_nonWSS} \\
A_{\rm C}^{\rm WSS}(t, \Delta t) &= \sum_{i=0}^{N} A_{\chi_{\rm R}}(\Delta t) A_{\chi_{\rm T}}(\Delta t)\notag\\
&\times\nbbE\left[ |\alpha_i|^2 {\rm e}^{-{\rm j} \frac{2\pi}{\lambda} y_{\rm D}\cos(\omega_i)\left[\theta(t + \Delta t) - \theta(t)\right]} \right]. \label{AC_t_WSS} 
\end{align}
Furthermore, when $\theta(t)$ is a random process with stationary increments, we have
\begin{align}
A_{\rm C}^{\rm WSS-SI}(\Delta t) &= \sum_{i=0}^{N} A_{\chi_{\rm R}}(\Delta t) A_{\chi_{\rm T}}(\Delta t) \notag\\
&\times\nbbE\left[ |\alpha_i|^2 {\rm e}^{-{\rm j} \frac{2\pi}{\lambda} y_{\rm D}\cos(\omega_i)\theta(\Delta t)} \right]. \label{AC_t_WSS_SI} 
\end{align}
\end{theorem}
\begin{IEEEproof}
Setting $\Delta f = 0$ in \eqref{F_ACF_nonWSS}, \eqref{F_ACF_WSS}, and \eqref{F_ACF_WSS_SI}, we end up with the final results.
\end{IEEEproof}
\begin{remark}\label{rem_no_imp}
In an ideal scenario where there is no hardware impairments and the UAV platform is ``completely" stable, we have $\theta(t) \equiv 0, ~ \forall t$, which results in a constant value for the channel ACF for all $t$ and $\Delta t$. Therefore, the coherence time will be infinity in this scenario.
\end{remark}


\begin{theorem} \label{thm3}
The coherence bandwidth of the air-to-ground channel can be formulated as in \eqref{CB}, where the ACFs for nonstationary and WSS hardware impairments are given as
\begin{align}
&A_{\rm C}^{\rm NonSt}\!(\Delta f; t) \!=\!\sum_{i=0}^{N} \!P_{\alpha_i}\! P_{\chi_{\rm R}(t)}\! \int_{0}^{\infty}\!\!\! P_{\chi_{\rm T}(t - \tau)} f_{\tau_i}(\tau) {\rm e}^{-{\rm j}2\pi \Delta f \tau}{\rm d}\tau, \label{AC_f_nonWSS} \\
&A_{\rm C}^{\rm WSS}(\Delta f) \!=\! \sum_{i=0}^{N} \!P_{\alpha_i}\! P_{\chi_{\rm R}}\! P_{\chi_{\rm T}}\! \int_{0}^{\infty} \!\!\!f_{\tau_i}(\tau) {\rm e}^{-{\rm j}2\pi \Delta f \tau}{\rm d}\tau, \label{AC_f_WSS}
\end{align}
respectively. Clearly, we also have $A_{\rm C}^{\rm WSS-SI}(\Delta f) = A_{\rm C}^{\rm WSS}(\Delta f)$. 
\end{theorem}
\begin{IEEEproof}
Setting $\Delta t = 0$ in \eqref{F_ACF_nonWSS} and \eqref{F_ACF_WSS}, we end up with the final results.
\end{IEEEproof}

\begin{remark} \label{remark1}
Since only the normalized values of the ACFs matter for the sake of evaluating the coherence bandwidth, we can ignore $P_{\chi_{\rm R}(t)}$ and $P_{\chi_{\rm R}} P_{\chi_{\rm T}}$ in \eqref{AC_f_nonWSS} and \eqref{AC_f_WSS}, respectively, as they are not functions of $\Delta f$ or the summation dummy variable $i$.
\end{remark}
\begin{remark} \label{remark2}
It is interesting to note that hardware impairments, when assumed to be WSS, as well as UAV wobbling do not have any impact on the coherence bandwidth of the channel. Furthermore, coherence bandwidth in the ideal no-impairment scenario will be the same as that of the WSS hardware impairments.
\end{remark}

Up until now, our focus was mainly on the characteristics of the channel, where using the channel impulse response $c(\tau; t)$, we studied the PDP, coherence time, and coherence bandwidth. Note that one can easily obtain other channel-related metrics, such as delay spread and Doppler spread by using the metrics derived in this section. Let us now study the behavior of the effective received noise at the ground UE by analyzing the PSD of the aggregate distortion caused by UAV wobbling and hardware impairments at both the transmitter and receiver plus the AWGN at the receiver, i.e., $n(t)$ defined in \eqref{n_HI}. The following theorem establishes this result.

\begin{theorem} \label{thm4}
The PSD of the distortion-plus-noise process $n(t)$ can be written as
\begin{align} \label{PSD_n_nonWSS}
&S_n^{\rm NonSt}(f) = \frac{N_0}{2} + \ncalF_{\Delta t}\{\langle A_{\eta_{\rm R}}(t, t+\Delta t)\rangle_t\}\, + \notag \\
& \sum_{i=0}^{N} \!\ncalF_{\Delta t}\bigg\{\!\Big\langle\! A_{\chi_{\rm R}}\!(t, t\!+\!\Delta t) \nbbE\left[|\alpha_i|^2 {\rm e}^{-{\rm j}\frac{2\pi}{\lambda} y_{\rm D}\cos(\omega_i)\left[\theta(t + \Delta t) - \theta(t)\right]}\right]\notag \\
&\quad\times  \int_0^\infty A_{\eta_{\rm T}}(t - \tau_i, t - \tau_i + \Delta t) f_{\tau_i}(\tau_i)\,{\rm d}\tau_i\Big\rangle_t\bigg\},
\end{align}
where $\frac{N_0}{2}$ is the AWGN power and $\langle .\rangle_t$ denotes averaging over time.
\end{theorem}
\begin{IEEEproof}
We start by writing the ACF of $n(t)$ as
\begin{align}
&A_n^{\rm NonSt}(t, t \!+\! \Delta t) = \nbbE\left[n^*(t)n(t\!+\!\Delta t)\right] \overset{(a)}{=} \nbbE\left[ \tilde{n}^*(t) \tilde{n}(t\!+\!\Delta t) \right]\notag \\
& + \nbbE\left[ \eta_{\rm R}^*(t) \eta_{\rm R}(t\!+\!\Delta t) \right] \!+\!\nbbE\Bigg[\!\bigg(\!\sum_{i=0}^{N}\! \alpha_i^* {\rm e}^{{\rm j}\varphi_i(t)} \chi_{\rm R}^*(t) \eta_{\rm T}^*(t \!-\! \tau_i(t))\bigg) \notag \\
&\times\! \bigg(\sum_{i=0}^{N} \alpha_i {\rm e}^{-{\rm j}\varphi_i(t+\Delta t)} \chi_{\rm R}(t\!+\!\Delta t) \eta_{\rm T}(t\!+\!\Delta t \!-\! \tau_i(t\!+\!\Delta t))\bigg)\Bigg] \notag \\
&\overset{(b)}{=} \!A_{\tilde{n}}(\Delta t) \!+\! A_{\eta_{\rm R}}(t, t\!+\!\Delta t) \!+\!\nbbE\bigg[\sum_{i=0}^{N} |\alpha_i|^2 {\rm e}^{-{\rm j}(\varphi_i(t+\Delta t)-\varphi_i(t))} \notag \\
&\quad\times \chi_{\rm R}^*(t)\chi_{\rm R}(t+\Delta t) \eta_{\rm T}^*(t - \tau_i) \eta_{\rm T}(t - \tau_i + \Delta t )\bigg] \notag\\
&= A_{\tilde{n}}(\Delta t) + A_{\eta_{\rm R}}(t, t+\Delta t) + \sum_{i=0}^{N} A_{\chi_{\rm R}}(t, t+\Delta t) \notag \\
&\quad\times \nbbE\left[|\alpha_i|^2 {\rm e}^{-{\rm j}\frac{2\pi}{\lambda} y_{\rm D}\cos(\omega_i)\left[\theta(t + \Delta t) - \theta(t)\right]}\right]\notag \\
&\quad\times \int_0^\infty A_{\eta_{\rm T}}(t - \tau_i, t - \tau_i + \Delta t) f_{\tau_i}(\tau_i)\,{\rm d}\tau_i, \notag
\end{align}
where in $(a)$ we used the independence between distortion noises and AWGN and the assumption that $\eta_{\rm R}(t)$ and $\tilde{n}(t)$ have zero means, and in $(b)$ we used the same reasoning as in the proof of Lemma \ref{lem:ACF} to demonstrate that the cross terms in the double-summation are zero, and we also used the approximation $\tau_i(t) \approx \tau_i(0) = \tau_i$. Note that $\tilde{n}(t)$ is a white process, and thus, we have $A_{\tilde{n}}(\Delta t) = \frac{N_0}{2}\delta(\Delta t)$, where $\frac{N_0}{2}$ is the noise power. To obtain the PSD of any nonstationary random process, we need to first average its ACF over time $t$, and then take its Fourier transform with respect to $\Delta t$ \cite[Sec. 7.2]{B_Peebels_Probability_2001}. Therefore, the PSD of $n(t)$ is written as in \eqref{PSD_n_nonWSS}.
\end{IEEEproof}

\begin{cor} \label{cor3}
The PSD of the distortion-plus-noise process $n(t)$ when hardware impairments are WSS can be written as
\begin{align} \label{PSD_n_WSS}
&S_n^{\rm WSS}(f) = \frac{N_0}{2} + \ncalF_{\Delta t}\{A_{\eta_{\rm R}}(\Delta t)\}+ \sum_{i=0}^{N} \ncalF_{\Delta t}\bigg\{ A_{\chi_{\rm R}}(\Delta t)\notag \\
& \times A_{\eta_{\rm T}}(\Delta t)\Big\langle \nbbE\left[|\alpha_i|^2 {\rm e}^{-{\rm j}\frac{2\pi}{\lambda} y_{\rm D}\cos(\omega_i)\left[\theta(t + \Delta t) - \theta(t)\right]}\right]\! \Big\rangle_t\!\bigg\}.
\end{align}
Also, assuming that $\theta(t)$ is a random process with stationary increments, this result is further simplified as
\begin{align} \label{PSD_n_WSS_SI}
&S_n^{\rm WSS-SI}(f) = \frac{N_0}{2} + \ncalF_{\Delta t}\{A_{\eta_{\rm R}}(\Delta t)\} + \sum_{i=0}^{N} \ncalF_{\Delta t}\bigg\{ A_{\chi_{\rm R}}(\Delta t) \notag \\
&\times A_{\eta_{\rm T}}(\Delta t) \nbbE\left[|\alpha_i|^2 {\rm e}^{-{\rm j}\frac{2\pi}{\lambda} y_{\rm D} \cos(\omega_i) \theta(\Delta t)}\right]\bigg\}.
\end{align}
\end{cor}

\begin{IEEEproof}
Following the proof of Theorem \ref{thm4} and assuming WSS hardware impairments, the ACF of $n(t)$ is simplified as
\begin{align}
&A_n^{\rm WSS}(t, t + \Delta t) = A_{\tilde{n}}(\Delta t) + A_{\eta_{\rm R}}(\Delta t)\notag\\
&+\! \sum_{i=0}^{N}\! A_{\chi_{\rm R}}(\Delta t) A_{\eta_{\rm T}}(\Delta t) \nbbE\!\left[|\alpha_i|^2 {\rm e}^{-{\rm j}\frac{2\pi}{\lambda} y_{\rm D}\!\cos(\omega_i)\left[\theta(t + \Delta t) - \theta(t)\right]}\right]\!, \notag
\end{align}
which results in \eqref{PSD_n_WSS} after averaging over $t$ and taking the Fourier transform with respect to $\Delta t$. Moreover, if we assume that $\theta(t)$ is a random process with stationary increments, the process $n(t)$ becomes stationary and we have
\begin{align}
&A_n^{\rm WSS-SI}(\Delta t) = A_{\tilde{n}}(\Delta t) + A_{\eta_{\rm R}}(\Delta t)\notag\\
&+ \sum_{i=0}^{N} A_{\chi_{\rm R}}(\Delta t) A_{\eta_{\rm T}}(\Delta t) \nbbE\left[|\alpha_i|^2 {\rm e}^{-{\rm j}\frac{2\pi}{\lambda} y_{\rm D}\cos(\omega_i)\theta(\Delta t)}\right], \notag
\end{align}
which gives \eqref{PSD_n_WSS_SI} after taking its Fourier transform.
\end{IEEEproof}
\begin{remark}
In the ideal no-impairment scenario (also see Remark \ref{rem_no_imp}) the PSD of $n(t)$ can be simplified as follows:
\begin{align*}
S_n^{\rm NoImp}(f) &= \frac{N_0}{2} + \left(P_{\eta_{\rm R}} + N P_{\chi_{\rm R}}P_{\eta_{\rm T}}P_{\alpha_i}\right)\delta(f),
\end{align*}
where the superscript ${\rm NoImp}$ stands for ``No-Impairment", $P_{\eta_{\rm R}} \coloneqq A_{\eta_{\rm R}}(0)$, and $P_{\eta_{\rm T}} \coloneqq A_{\eta_{\rm T}}(0)$.
\end{remark}

\subsection{Metrics: Evaluations for Some Case Studies} \label{subsec:Eval}
The results we have derived so far are general and hold for any UAV wobbling model and any distribution of delay and distortion noise processes. However, to gain useful insights, it is instructive to simplify these results for specific case studies. We consider some reasonable models in this section and evaluate the metrics derived in the previous section accordingly.

\subsubsection{UAV Wobbling} \label{subsubsec:Wobble}
We consider two different random processes for modeling the UAV pitch angle $\theta(t)$: (i) Wiener process, and (ii) sinusoidal process. Let us review the properties of each process first.
\begin{itemize}
\item \textbf{Stationary-Increments -- Wiener Process}: This canonical model is defined as a process with independent, stationary, and Gaussian increments that is continuous in time $t$. Furthermore, its value at $t = 0$ is assumed to be zero. Therefore, we have $\theta(t + \Delta t) - \theta(t) \overset{{\rm dist.}}{=\joinrel=} \theta(\Delta t) \sim \ncalN\left(0, \Delta t\right)$, where $\overset{{\rm dist.}}{=\joinrel=}$ stands for equality in distribution. Since Wiener processes are scale-invariant, we can make the variance of $\theta(\Delta t)$ dimensionless by multiplying it with the proportionality constant $b = 1 {\rm ~s^{-1}}$. Following the characteristic function (cf) of Gaussian random variables, we have
\begin{align} \label{cf_win}
\nbbE\left[{\rm e}^{{\rm j}\omega (\theta(t + \Delta t) - \theta(t))}\right] = \nbbE\left[{\rm e}^{{\rm j}\omega \theta(\Delta t)}\right] = {\rm e}^{-\frac{\omega^2}{2} \Delta t}.
\end{align}

\item \textbf{Non-Stationary-Increments -- Sinusoidal Process}: This model is motivated by the random \textit{oscillatory} behavior of wobbling. We assume that $\theta(t) = L \sin(2\pi Q t)$, where $L \sim U[-\theta_{\rm m}, \theta_{\rm m})$ and $Q \sim f_Q(q)$ are the amplitude and frequency of the variations of the UAV pitch angle, respectively, with $\theta_{\rm m}$ and $f_Q(.)$ being the maximum UAV pitch angle and some given pdf, respectively. We also assume that $L$ and $Q$ are independent. Clearly, the independent and stationary-increments property does not hold for this process. Following the cf of uniform random variables, we have $\nbbE\left[{\rm e}^{{\rm j}\omega (\theta(t + \Delta t) - \theta(t))}\right] =$
\begin{align} \label{cf_sin}
\int_{-\infty}^{\infty}\!\!\!\sinc\left(\frac{\omega}{\pi}\theta_{\rm m}\left(\sin(2\pi q (t\!+\!\Delta t))\!-\!\sin(2\pi q t)\right)\right)\!f_Q(q){\rm d}q,
\end{align}
where $\sinc(x) = \frac{\sin(\pi x)}{\pi x}$.
\end{itemize}
%

Using \eqref{Laplace}, \eqref{cf_win}, and \eqref{cf_sin}, we introduce and evaluate the following relations which will be used extensively in our derivations:
\begin{align}
	&G_i^{\rm SI}(\Delta t) = \nbbE\left[ |\alpha_i|^2 {\rm e}^{-{\rm j} \frac{2\pi}{\lambda} y_{\rm D}\cos(\omega_i)\theta(\Delta t)} \right] \notag\\
	&=
	\begin{cases}
		\int_0^{\pi/2} \frac{1}{\pi \beta} {\rm e}^{-\frac{|\omega_i - \omega_0|}{\beta}} {\rm e}^{-\frac{2\pi^2}{\lambda^2}y_{\rm D}^2\cos^2(\omega_i)\Delta t} \,{\rm d}\omega_i, &\! i \neq 0 \\
		2NKP_{\alpha_1} {\rm e}^{-\frac{2\pi^2}{\lambda^2}y_{\rm D}^2\cos^2(\omega_0)\Delta t}, &\! i = 0,
	\end{cases} \label{Gi_SI} \\
	&G_i^{\rm NonSI}(t, \Delta t) = \nbbE\left[ |\alpha_i|^2 {\rm e}^{-{\rm j} \frac{2\pi}{\lambda} y_{\rm D}\cos(\omega_i)\left[\theta(t + \Delta t) - \theta(t)\right]} \right] \notag \\
	&=
	\begin{cases}
		\int_0^{\pi/2}\int_{-\infty}^{\infty} \frac{1}{\pi \beta} {\rm e}^{-\frac{|\omega_i - \omega_0|}{\beta}} f_Q(q)\sinc\big(\frac{2}{\lambda} \theta_{\rm m} y_{\rm D} \cos(\omega_i) \\
		\quad\times\left(\sin(2\pi q (t+\Delta t))-\sin(2\pi q t)\right)\!\big){\rm d}q\,{\rm d}\omega_i, &\hspace{-0.5cm} i \neq 0 \\
		2NKP_{\alpha_1}\int_{-\infty}^{\infty} f_Q(q)\sinc\big(\frac{2}{\lambda} \theta_{\rm m} y_{\rm D} \cos(\omega_0)  \\
		\quad\times\left(\sin(2\pi q (t+\Delta t))-\sin(2\pi q t)\right)\!\big) {\rm d}q, &\hspace{-0.5cm} i = 0,
	\end{cases} \label{Gi}
\end{align}
where the superscript ${\rm NonSI}$ stands for ``Non-Stationary-Increments". Note that the functions $G_i^{\rm NonSI}(t, \Delta t), ~ 1 \leq i \leq N$ are all equal to each other. The same is also true for the functions $G_i^{\rm SI}(\Delta t), ~ 1 \leq i \leq N$.

\begin{remark}
The Wiener process is proposed as a purely theoretical model that provides a reasonable baseline for the more practical sinusoidal process, where the latter is a realistic model that captures both oscillatory behavior and randomness of UAV wobbling. Note that the sinusoidal process does not have stationary increments because of which the received signal is nonstationary (the channel ACF will be a function of both $t$ and $\Delta t$).
\end{remark}

\subsubsection{Distortion Noises} \label{subsubsec:Distortion}
As for the additive distortion noise, we assume WSS Gaussian processes both at the transmitter and receiver, which is backed by measurements and theoretical studies \cite{B_Schenk_Imperfections_2008, J_Emil_New_2013}. Although various models are available for the covariance function of Gaussian processes \cite{B_Rasmussen_Gaussian_2006}, we consider the squared exponential model in this paper. Hence, the ACFs for the additive distortion noises can be written as
\begin{align} \label{A_eta}
A_{\eta_{\rm T}}(\Delta t) = \kappa_{\eta_{\rm T}}^2 {\rm e}^{-\frac{(\Delta t)^2}{2l_{\eta_{\rm T}}^2}}, \qquad A_{\eta_{\rm R}}(\Delta t) = \kappa_{\eta_{\rm R}}^2 {\rm e}^{-\frac{(\Delta t)^2}{2l_{\eta_{\rm R}}^2}},
\end{align}
where $\kappa_{\eta_{\rm T}}^2$ ($\kappa_{\eta_{\rm R}}^2$) and $l_{\eta_{\rm T}}$ ($l_{\eta_{\rm R}}$) are design parameters, which represent the maximum power and the characteristic length-scale of the additive distortion noise process at the transmitter (receiver), respectively. On the other hand, the research on representative models for the multiplicative distortion noise processes is scarce. As also pointed out in \cite[p. 261]{B_Schenk_Imperfections_2008}, researchers usually ignore the time-dependence and randomness of these distortion processes and assume deterministic multiplicative errors. In this paper, we study two representative models based on random processes (WSS and nonstationary) for these multiplicative errors listed below:
\begin{itemize}
\item \textbf{WSS -- Gaussian Process}: Similar to the additive distortion noise, we assume the canonical Gaussian process for the WSS multiplicative distortion noises. The ACFs in this case can be written as
\begin{align}
A_{\chi_{\rm T}}(\Delta t) = \kappa_{\chi_{\rm T}}^2 {\rm e}^{-\frac{(\Delta t)^2}{2l_{\chi_{\rm T}}^2}}, \qquad A_{\chi_{\rm R}}(\Delta t) = \kappa_{\chi_{\rm R}}^2 {\rm e}^{-\frac{(\Delta t)^2}{2l_{\chi_{\rm R}}^2}} \label{A_chi_WSS},
\end{align}
where $\kappa_{\chi_{\rm T}}^2$, $\kappa_{\chi_{\rm R}}^2$, $l_{\chi_{\rm T}}$, and $l_{\chi_{\rm R}}$ are defined similar as before.
\item \textbf{Nonstationary -- Sinusoidal Process}: As a canonical nonstationary model, we again use the sinusoidal process in this case that was also used to model UAV wobbling above. In particular, we define $\chi_{\rm T}(t) = L_{\rm T}\sin(2\pi Q_{\rm T} t)$ and $\chi_{\rm R}(t) = L_{\rm R}\sin(2\pi Q_{\rm R} t)$, where $L_{\rm T}$ ($L_{\rm R}$) and $Q_{\rm T}$ ($Q_{\rm R}$) are the amplitude and frequency of variations of $\chi_{\rm T}(t)$ ($\chi_{\rm R}(t)$), respectively. Therefore, we can write their ACFs as
\begin{align}
&A_{\chi_{\rm T}}(t, t + \Delta t) =\notag\\
&\,\,P_{L_{\rm T}} \int_{-\infty}^{\infty} \sin(2\pi q t) \sin(2\pi q (t + \Delta t)) f_{Q_{\rm T}}(q)\,{\rm d}q, \label{A_chi_T_nonWSS} \\
&A_{\chi_{\rm R}}(t, t + \Delta t) =\notag\\
&\,\,P_{L_{\rm R}} \int_{-\infty}^{\infty} \sin(2\pi q t) \sin(2\pi q (t + \Delta t)) f_{Q_{\rm R}}(q)\,{\rm d}q, \label{A_chi_R_nonWSS}
\end{align}
where $P_{L_{\rm T}} = \nbbE\left[|L_{\rm T}|^2\right]$, $P_{L_{\rm R}} = \nbbE\left[|L_{\rm R}|^2\right]$, and $f_{Q_{\rm T}}(.)$ and $f_{Q_{\rm R}}(.)$ are the pdfs of $Q_{\rm T}$ and $Q_{\rm R}$, respectively. Further discussion about the choice of different parameters for this model will be provided in Section \ref{sec:simulation}.
\end{itemize}

\begin{remark} \label{remark3}
In the nonstationary setup, one could also assume a Wiener process for the multiplicative distortion noise processes. However, since the ACF of the Wiener process $W(t)$ is simply the minimum value of its observation times, i.e., $A_W(t, t + \Delta t) = t$ for non-negative $\Delta t$, all of our metrics will quickly diverge using this model. Hence, the Wiener process is not a suitable model for the multiplicative distortion noise.
\end{remark}

\begin{remark}
As mentioned earlier, RF impairments are nonstationary in nature and represent time-dependent behaviors. We extend the generalized error model proposed in \cite[Ch. 7]{B_Schenk_Imperfections_2008} by statistical characterization of the additive and multiplicative terms (as opposed to deterministic terms), which are more suitable for wireless systems with RF imperfections, such as the aerial-terrestrial setup considered in this paper.
\end{remark}

Using these models for UAV wobbling and distortion noise processes, we will now evaluate the metrics derived in Section \ref{subsec:General}.

\textbf{\textit{PDP}}: Using Theorem \ref{thm1}, we obtain the PDP for nonstationary and WSS hardware impairments, respectively, as
\begin{align}
&P_{\rm c}^{\rm NonSt}(\tau; t) = P_{\alpha_1} P_{L_{\rm R}}P_{L_{\rm T}} \int_{-\infty}^{\infty} \sin^2(2\pi q t) f_{Q_{\rm R}}(q)\,{\rm d}q \notag \\
&\quad\times \int_{-\infty}^{\infty} \sin^2(2\pi q (t - \tau)) f_{Q_{\rm T}}(q)\,{\rm d}q \notag\\
&\quad\times\left(NK \delta(\tau - \tau_0) + \sum_{i=1}^{N}  \rho_i{\rm e}^{-\rho_i(\tau - \tau_0)} {\bf 1}(\tau - \tau_0)\right), \label{PDP_nonWSS_1} \\
&P_{\rm c}^{\rm WSS}(\tau) = P_{\alpha_1} \kappa_{\chi_{\rm R}}^2 \kappa_{\chi_{\rm T}}^2\notag\\
&\quad\times\left(NK \delta(\tau - \tau_0) + \sum_{i=1}^{N}  \rho_i{\rm e}^{-\rho_i(\tau - \tau_0)} {\bf 1}(\tau - \tau_0)\right), \label{PDP_WSS_1}
\end{align}
where we used the relations $P_{\alpha_i} = P_{\alpha_1} = \frac{1}{2\pi}\big( 2 - {\rm e}^{-\frac{\omega_0}{\beta}} - {\rm e}^{-\frac{\pi/2 - \omega_0}{\beta}} \big), ~ 1 \leq i \leq N$, and $P_{\alpha_0} = NKP_{\alpha_1}$. From \eqref{mu-WSS} and \eqref{sigma-WSS}, we obtain the average and rms delay spreads for WSS hardware impairments, respectively, as
\begin{align}
\mu^{\rm WSS} &= \tau_0 + \frac{1}{N(K+1)}\sum_{i = 1}^{N}\frac{1}{\rho_i}, \label{mu_WSS_1} \\
\sigma^{\rm WSS} &= \sqrt{\frac{2}{N(K\!+\!1)}\!\sum_{i = 1}^{N}\!\frac{1}{\rho_i^2} \!-\! \frac{1}{N^2(K\!+\!1)^2}\!\left(\sum_{i = 1}^{N}\frac{1}{\rho_i}\right)^{\!\!2}}. \label{sigma_WSS_1}
\end{align}

\textbf{\textit{Coherence Time}}: From Theorem \ref{thm2}, \eqref{Gi_SI}, and \eqref{Gi}, one can obtain the coherence time of the channel for different impairment scenarios. Since the equations become quite intricate in all cases, we need to perform numerical integration to calculate the coherence time. However, when hardware impairments are WSS, the final equations could be simplified further. Using the predefined functions $G_i^{\rm NonSI}(t, \Delta t)$ and $G_i^{\rm SI}(\Delta t)$, we can simplify the ACFs as
\begin{align*}
A_{\rm C}^{\rm WSS}(t, \Delta t) &= \kappa_{\chi_{\rm R}}^2 \kappa_{\chi_{\rm T}}^2 {\rm e}^{-\left(\frac{1}{2l_{\chi_{\rm R}}^2} + \frac{1}{2l_{\chi_{\rm T}}^2}\right)(\Delta t)^2}\\
&\quad\times\left(G_0^{\rm NonSI}(t, \Delta t) + NG_1^{\rm NonSI}(t, \Delta t)\right),\\
A_{\rm C}^{\rm WSS-SI}(\Delta t) &= \kappa_{\chi_{\rm R}}^2 \kappa_{\chi_{\rm T}}^2 {\rm e}^{-\left(\frac{1}{2l_{\chi_{\rm R}}^2} + \frac{1}{2l_{\chi_{\rm T}}^2}\right)(\Delta t)^2}\\
&\quad\times\left(G_0^{\rm SI}(\Delta t) + NG_1^{\rm SI}(\Delta t)\right),
\end{align*}
where we used the fact that $G_i^{\rm NonSI}(t, \Delta t) = G_1^{\rm NonSI}(t, \Delta t), ~1 \leq i \leq N$, and $G_i^{\rm SI}(\Delta t) = G_1^{\rm SI}(\Delta t), ~1 \leq i \leq N$. Note that the maximum value for both ACFs is $2 \kappa_{\chi_{\rm R}}^2 \kappa_{\chi_{\rm T}}^2 N(K+1)P_{\alpha_1}$, which occurs at $\Delta t = 0$. Therefore, the coherence time for WSS hardware impairments when $\theta(t)$ does not and does have the stationary-increments property can be written, respectively, as
\begin{align}
&T_{\rm Coh}^{\rm WSS}(t) = \min\bigg\{\Delta t: \frac{{\rm e}^{-\big(\frac{1}{2l_{\chi_{\rm R}}^2} + \frac{1}{2l_{\chi_{\rm T}}^2}\big)(\Delta t)^2}}{2N(K+1)P_{\alpha_1}}\notag\\
&\qquad\times \left|G_0^{\rm NonSI}(t, \Delta t) + NG_1^{\rm NonSI}(t, \Delta t)\right| \leq \gamma_{\rm T}\bigg\}, \label{T_Coh_WSS}\\
&T_{\rm Coh}^{\rm WSS-SI} = \min\bigg\{\Delta t: \frac{{\rm e}^{-\big(\frac{1}{2l_{\chi_{\rm R}}^2} + \frac{1}{2l_{\chi_{\rm T}}^2}\big)(\Delta t)^2}}{2N(K+1)P_{\alpha_1}}\notag\\
&\qquad\times \left|G_0^{\rm SI}(\Delta t) + NG_1^{\rm SI}(\Delta t)\right| \leq \gamma_{\rm T}\bigg\} \label{T_Coh_WSS_SI},
\end{align}
which needs to be solved numerically for $\Delta t$. These equations can be further simplified for the limiting case of $K \to \infty$, i.e., a purely LoS channel with no multipath, as follows:
\begin{align*}
&T_{\rm Coh}^{\rm WSS, \,LoS}(t) \!=\!\! \lim_{K\to \infty}\!\! T_{\rm Coh}^{\rm WSS}(t) \!=\! \min\!\Bigg\{\!\Delta t\!:\! {\rm e}^{\!-\big(\frac{1}{2l_{\chi_{\rm R}}^2} + \frac{1}{2l_{\chi_{\rm T}}^2}\big)(\Delta t)^2} \\
&\quad\times\int_{-\infty}^{\infty} f_Q(q)\sinc\bigg(\frac{2}{\lambda} \theta_{\rm m} y_{\rm D} \cos(\omega_0) \notag\\
&\quad\times\left(\sin(2\pi q (t+\Delta t))-\sin(2\pi q t)\right)\!\bigg) {\rm d}q \leq \gamma_{\rm T}\Bigg\}, \\
&T_{\rm Coh}^{\rm WSS-SI, \,LoS} = \lim_{K\to \infty} T_{\rm Coh}^{\rm WSS-SI} \\
&= \min\left\{\Delta t\!:\! {\rm e}^{-\big(\frac{1}{2l_{\chi_{\rm R}}^2} + \frac{1}{2l_{\chi_{\rm T}}^2}\big)(\Delta t)^2}{\rm e}^{-\frac{2\pi^2}{\lambda^2}y_{\rm D}^2\cos^2(\omega_0)\Delta t} \leq \gamma_{\rm T}\right\}\\
&=\scalebox{1.1}{$\frac{\sqrt{\frac{4\pi^4}{\lambda^4}y_{\rm D}^4\cos^4(\omega_0) - 2\log(\gamma_{\rm T})\big(\frac{1}{l_{\chi_{\rm R}}^2} + \frac{1}{l_{\chi_{\rm T}}^2}\big)} - \frac{2\pi^2}{\lambda^2}y_{\rm D}^2\cos^2(\omega_0)}{\frac{1}{l_{\chi_{\rm R}}^2} + \frac{1}{l_{\chi_{\rm T}}^2}}$}.
\end{align*}

\textbf{\textit{Coherence Bandwidth}}: We use Theorem \ref{thm3} along with \eqref{A_chi_WSS}, \eqref{A_chi_T_nonWSS}, and \eqref{A_chi_R_nonWSS} to derive the coherence bandwidth of the channel. Following Remark \ref{remark1}, we do not need to evaluate $P_{\chi_{\rm R}(t)}$ and $P_{\chi_{\rm R}} P_{\chi_{\rm T}}$ in \eqref{AC_f_nonWSS} and \eqref{AC_f_WSS}, respectively, as they will be canceled out when normalizing the ACF. Let us start with simplifying the channel ACF when hardware impairments are nonstationary. Setting $\Delta t = 0$ and $t \mapsto t - \tau$ in \eqref{A_chi_T_nonWSS}, and swapping the integrals in \eqref{AC_f_nonWSS}, we can write
\begin{align*}
&A_{\rm C}^{\rm NonSt}(\Delta f; t) = P_{\alpha_1} P_{\chi_{\rm R}(t)} P_{L_{\rm T}} \Bigg[NK \int_{-\infty}^{\infty} f_{Q_{\rm T}}(q)\\
&\times\!\! \int_{0}^{\infty} \!\!\!\!\sin^2(2\pi q (t \!-\! \tau)) \delta(\tau \!-\! \tau_0) {\rm e}^{-{\rm j}2\pi \Delta f \tau} {\rm d}\tau{\rm d}q \!+ \!\!\sum_{i=1}^{N} \!\int_{-\infty}^{\infty}\!\!\!\! f_{Q_{\rm T}}(q) \\
& \times \int_{\tau_0}^{\infty}  \sin^2(2\pi q (t - \tau)) \rho_i{\rm e}^{-\rho_i(\tau - \tau_0)} {\rm e}^{-{\rm j}2\pi \Delta f \tau} \,{\rm d}\tau \,{\rm d}q \Bigg]\\
&= P_{\alpha_1} P_{\chi_{\rm R}(t)} P_{L_{\rm T}} {\rm e}^{-{\rm j}2\pi \Delta f \tau_0} \left[H_0(t) + \sum_{i=1}^{N} H_i(\Delta f; t) \right],
\end{align*}
where
\begin{align*}
&H_0(t) = NK \int_{-\infty}^{\infty} f_{Q_{\rm T}}(q) \sin^2(2\pi q (t - \tau_0)) \,{\rm d}q, \\
&H_i(\Delta f; t) = \scalebox{1.3}{$\frac{\rho_i}{2(\rho_i + {\rm j}2\pi\Delta f)}$} - \int_{-\infty}^{\infty} f_{Q_{\rm T}}(q) \\
&\times\left(\scalebox{1.3}{$\frac{\rho_i{\rm e}^{{\rm j}4\pi q (t - \tau_0)}}{4(\rho_i + {\rm j}2\pi\Delta f + {\rm j}4\pi q)} + \frac{\rho_i{\rm e}^{-{\rm j}4\pi q (t - \tau_0)}}{4(\rho_i + {\rm j}2\pi\Delta f - {\rm j}4\pi q)} $}\right) {\rm d}q.
\end{align*}
Taking the absolute value of $A_{\rm C}^{\rm NonSt}(\Delta f; t)$, we can write the coherence bandwidth as
\begin{align}
B_{\rm Coh}^{\rm NonSt}\!(t) \!=\! \min\!\left\{\!\!\Delta f\!:\!\! \frac{\left|H_0(t) \!+\! \sum_{i=1}^{N}\! H_i(\Delta f; t) \right|}{\max{\left|H_0(t) \!+\! \sum_{i=1}^{N}\! H_i(\Delta f; t) \right|}} \!\leq\! \gamma_{\rm B}\!\right\}\!, \label{B_Coh_nonWSS}
\end{align}
which needs to be solved numerically for $\Delta f$. On the other hand, for WSS hardware impairments, the channel ACF can be further simplified as
\begin{align*}
A_{\rm C}^{\rm WSS}(\Delta f) &= \!P_{\alpha_1} P_{\chi_{\rm R}} P_{\chi_{\rm T}} \Bigg[NK \int_{0}^{\infty} \delta(\tau - \tau_0) {\rm e}^{-{\rm j}2\pi \Delta f \tau} \,{\rm d}\tau\\
&\quad + \sum_{i=1}^{N} \int_{\tau_0}^{\infty} \rho_i{\rm e}^{-\rho_i(\tau - \tau_0)} {\rm e}^{-{\rm j}2\pi \Delta f \tau} \,{\rm d}\tau \Bigg]\\
&= \!P_{\alpha_1} \!P_{\chi_{\rm R}} \!P_{\chi_{\rm T}} {\rm e}^{-{\rm j}2\pi \Delta f \tau_0} \!\!\left[\!NK \!+\! \sum_{i=1}^{N}\! \frac{\rho_i}{\rho_i \!+\! {\rm j}2\pi\Delta f} \!\right]\!,
\end{align*}
which gives the coherence bandwidth as
\begin{align}
B_{\rm Coh}^{\rm WSS} = \min\left\{\!\Delta f\!:\! \frac{\left|K \!+\! \frac{1}{N}\sum_{i=1}^{N} \frac{\rho_i}{\rho_i + {\rm j}2\pi\Delta f} \right|}{K\!+\!1} \!\leq\! \gamma_{\rm B}\right\}\!. \label{B_Coh_WSS}
\end{align}
Observe that the coherence bandwidth is not a function of UAV wobbling or hardware impairments in this case (see Remark \ref{remark2}). Similar to the case of coherence time, for $K \to \infty$, one can easily see that the terms containing $H_i(\Delta f; t)$ can be ignored with respect to $H_0(t)$, which makes the solutions of \eqref{B_Coh_nonWSS} and \eqref{B_Coh_WSS} trivial. In fact, for any $\gamma_{\rm B} < 1$ in this case, we have $B_{\rm Coh}^{\rm NonSt, \,LoS}(t) = B_{\rm Coh}^{\rm WSS, \,LoS} = \infty$. This observation can be seen in Fig. \ref{fig:CB_2}, as well.

\textbf{\textit{PSD of Distortion-Plus-Noise}}: Theorem \ref{thm4} gives the PSD of the distortion-plus-noise process $n(t)$ for the general case where hardware impairments are nonstationary and UAV wobbling does not necessarily have the stationary-increments property. One could use \eqref{cf_sin}, \eqref{A_eta}, and \eqref{A_chi_R_nonWSS} to obtain an equation for the PSD of $n(t)$, but as is clear from the form of these equations, this derivation quickly becomes intractable. Therefore, for the sake of simplicity, we only analytically evaluate the PSD of the distortion-plus-noise process when hardware impairments are WSS. Let us first analyze the PSD of $n(t)$ when UAV wobbling has the stationary-increments property. Following \eqref{PSD_n_WSS_SI} in Corollary \ref{cor3}, we have $S_n^{\rm WSS-SI}(f)$
\begin{align} \label{PSD_n_WSS_SI_1}
&= \frac{N_0}{2} + \ncalF_{\Delta t}\left\{\kappa_{\eta_{\rm R}}^2 {\rm e}^{-\frac{(\Delta t)^2}{2l_{\eta_{\rm R}}^2}}\right\} \notag\\
&\quad+ \sum_{i=0}^{N} \ncalF_{\Delta t}\left\{ \kappa_{\chi_{\rm R}}^2 {\rm e}^{-\frac{(\Delta t)^2}{2l_{\chi_{\rm R}}^2}} \kappa_{\eta_{\rm T}}^2 {\rm e}^{-\frac{(\Delta t)^2}{2l_{\eta_{\rm T}}^2}} G_i^{\rm SI}(\Delta t)\right\}\notag \\
&= \frac{N_0}{2} + \ncalF_{\Delta t}\left\{\kappa_{\eta_{\rm R}}^2 {\rm e}^{-\frac{(\Delta t)^2}{2l_{\eta_{\rm R}}^2}}\right\}  \notag\\
&\quad+ \ncalF_{\Delta t}\left\{ \kappa^2 {\rm e}^{-\frac{(\Delta t)^2}{2l^2}} 2NKP_{\alpha_1} {\rm e}^{-u^2\cos^2(\omega_0)\Delta t} \right\}\notag \\
&\quad + N\ncalF_{\Delta t}\left\{\! \kappa^2 {\rm e}^{-\frac{(\Delta t)^2}{2l^2}} \!\!\!\int_0^{\pi/2}\!\!\! \frac{1}{\pi \beta} {\rm e}^{-\frac{|\omega_i - \omega_0|}{\beta}} {\rm e}^{-u^2\cos^2(\omega_i)\Delta t} {\rm d}\omega_i \!\right\}\notag \\
&= \frac{N_0}{2} + \kappa_{\eta_{\rm R}}^2 \sqrt{2\pi l_{\eta_{\rm R}}^2} {\rm e}^{-2\pi^2 l_{\eta_{\rm R}}^2 f^2}  \notag\\
&\quad+  \kappa^2 \sqrt{2\pi l^2} {\rm e}^{-2\pi^2 l^2 f^2} \bigg(\!2P_{\alpha_0} {\rm e}^{\frac{1}{2}l^2 u^4 \cos^4(\omega_0)} {\rm e}^{{\rm j}2\pi l^2 u^2 \cos^2(\omega_0) f}\notag \\
&\quad + N \!\!\int_0^{\pi/2} \!\!\!\frac{1}{\pi \beta} {\rm e}^{-\frac{|\omega_i - \omega_0|}{\beta}} \!{\rm e}^{\frac{1}{2}l^2 u^4 \cos^4(\omega_i)} {\rm e}^{{\rm j}2\pi l^2 u^2 \cos^2(\omega_i) f} {\rm d}\omega_i\bigg),
\end{align}
where we defined the parameters $\kappa^2 = \kappa_{\chi_{\rm R}}^2 \kappa_{\eta_{\rm T}}^2$, $l^{-2} = l_{\chi_{\rm R}}^{-2} + l_{\eta_{\rm T}}^{-2}$, and $u^2 = \frac{2\pi^2}{\lambda^2}y_{\rm D}^2$ to simplify the equations, and in the last equality we used the Fourier transform of a Gaussian function, i.e., $\ncalF_{x}\left\{ {\rm e}^{-ax^2} \right\} = \sqrt{\frac{\pi}{a}} {\rm e}^{-\frac{\pi^2}{a}f^2}$. As for the case where UAV wobbling does not have the stationary-increments property (see \eqref{PSD_n_WSS} in Corollary \ref{cor3}), we need to first average $G_i^{\rm NonSI}(t, \Delta t)$ over time $t$. Following the same procedure as above, we can write the PSD of $n(t)$ as $S_n^{\rm WSS}(f)$
\begin{align} \label{PSD_n_WSS_1}
&= \frac{N_0}{2} + \ncalF_{\Delta t}\left\{\kappa_{\eta_{\rm R}}^2 {\rm e}^{-\frac{(\Delta t)^2}{2l_{\eta_{\rm R}}^2}}\right\}  \notag\\
&\quad+ \sum_{i=0}^{N} \ncalF_{\Delta t}\left\{ \kappa_{\chi_{\rm R}}^2 {\rm e}^{-\frac{(\Delta t)^2}{2l_{\chi_{\rm R}}^2}} \kappa_{\eta_{\rm T}}^2 {\rm e}^{-\frac{(\Delta t)^2}{2l_{\eta_{\rm T}}^2}} \left\langle G_i^{\rm NonSI}(t, \Delta t) \right\rangle_t\right\}\notag \\
&= \frac{N_0}{2} + \kappa_{\eta_{\rm R}}^2 \sqrt{2\pi l_{\eta_{\rm R}}^2} {\rm e}^{-2\pi^2 l_{\eta_{\rm R}}^2 f^2}  \notag\\
&\quad+ 2P_{\alpha_0} \lim_{T\to\infty} \frac{1}{T} \int_{0}^{T} \int_{-\infty}^{\infty}  \int_{-\infty}^{\infty} \kappa^2 {\rm e}^{-\frac{(\Delta t)^2}{2l^2}} \notag\\
&\quad\times\sinc\left(\frac{2}{\lambda} \theta_{\rm m} y_{\rm D} \cos(\omega_0) \left(\sin(2\pi q (t\!+\!\Delta t))\!-\!\sin(2\pi q t)\right)\right) \notag\\
&\quad\times f_Q(q) {\rm e}^{-{\rm j}2\pi f \Delta t}\, {\rm d}q \,{\rm d}\Delta t \,{\rm d}t \notag\\
&\quad + N \lim_{T\to\infty} \frac{1}{T} \int_{0}^{T} \int_{-\infty}^{\infty} \int_{0}^{\frac{\pi}{2}} \int_{-\infty}^{\infty} \kappa^2 {\rm e}^{-\frac{(\Delta t)^2}{2l^2}} \frac{1}{\pi \beta} {\rm e}^{-\frac{|\omega_i - \omega_0|}{\beta}} \notag\\
&\quad\times\sinc\left(\frac{2}{\lambda} \theta_{\rm m} y_{\rm D} \cos(\omega_i) \left(\sin(2\pi q (t\!+\!\Delta t))\!-\!\sin(2\pi q t)\right)\!\right) \notag\\
&\quad\times f_Q(q) {\rm e}^{-{\rm j}2\pi f \Delta t} \,{\rm d}q \,{\rm d}\omega_i \,{\rm d}\Delta t \,{\rm d}t,
\end{align}
where in the last equality we used the definition of Fourier transform and swapped the order of limit and integral.
\section{Simulation Results} \label{sec:simulation}
In this section, we provide numerical results to obtain further insights on the impact of UAV wobbling and hardware impairments on the key performance metrics of wireless channels. We assume that a rotary-winged UAV hovers in the air and communicates with a UE on the ground in a multi-path channel. The parameters of the channel used in this paper are as follows: Number of MPCs $N \in \{20, 10\}$ for sub-$6$ GHz and mmWave frequencies, respectively, carrier frequency $f_{\rm c} \in \{2.4, 6, 30\}$ GHz (thus, signal wavelength $\lambda \in \{12.5, 5, 1\}$ cm), UAV-UE distance $d_0 = 300$ m (thus, UAV-UE delay $\tau_0 = 1~\mu$s), UAV-UE AoD $\omega_0 = 20^\circ$, reciprocal of the mean excess delay of the $i$-th MPC $\rho_i \sim U[10^7, 10^8)~{\rm s}^{-1}$, Rician $K$ factor $K \in \{0, 1, 5, 11.5\}$ \cite[Table 1]{C_Goddemeier_Investigation_2015} (note that $K = 0$ represents the Rayleigh channel), scaling parameter of the Laplacian angular power spectrum $\beta = 1$, and noise power $\frac{N_0}{2} = 5\times 10^{-9}$ W. The UAV antenna is placed at a distance of $y_{\rm D} = 40$ cm from the UAV platform centroid. Note that these are just nominal values and our results are not impacted by their choice. Due to many reasons discussed earlier, the UAV may wobble, which is modeled using two random processes, i.e., the Wiener and sinusoidal processes. The UAV pitch angle in the latter case is bounded to $\theta_{\rm m} \in \{5, 7, 10\}^\circ$ \cite{J_Ahmed_Flight_2010} with a frequency of variations that is uniformly distributed between $5$ and $25$ Hz. Both the UAV and the UE suffer from hardware impairments as well, and the parameters defined in this paper for modeling the hardware impairments are as follows: Maximum power and the characteristic length-scale of the WSS additive/multiplicative distortion noise processes at the transmitter/receiver are $\{\kappa_{\eta_{\rm T}}^2, \kappa_{\eta_{\rm R}}^2, \kappa_{\chi_{\rm T}}^2, \kappa_{\chi_{\rm R}}^2\} \in \{0.1, 0.5, 1\}$ W and $\{l_{\eta_{\rm T}}, l_{\eta_{\rm R}}, l_{\chi_{\rm T}}, l_{\chi_{\rm R}}\} \in \{0.01, 0.05, 0.1\}$ s, respectively, and power of amplitude and frequency of variations of nonstationary multiplicative distortion noise process at the transmitter/receiver are $P_{L_{\rm T}} = P_{L_{\rm R}} = 0.5$ W and $\{Q_{\rm T}, Q_{\rm R}\} \sim U[5, 15)$ Hz, respectively.

\begin{figure}[t!]
	\centering
	\begin{minipage}{0.49\columnwidth}
		\centering
		\includegraphics[width=1\textwidth]{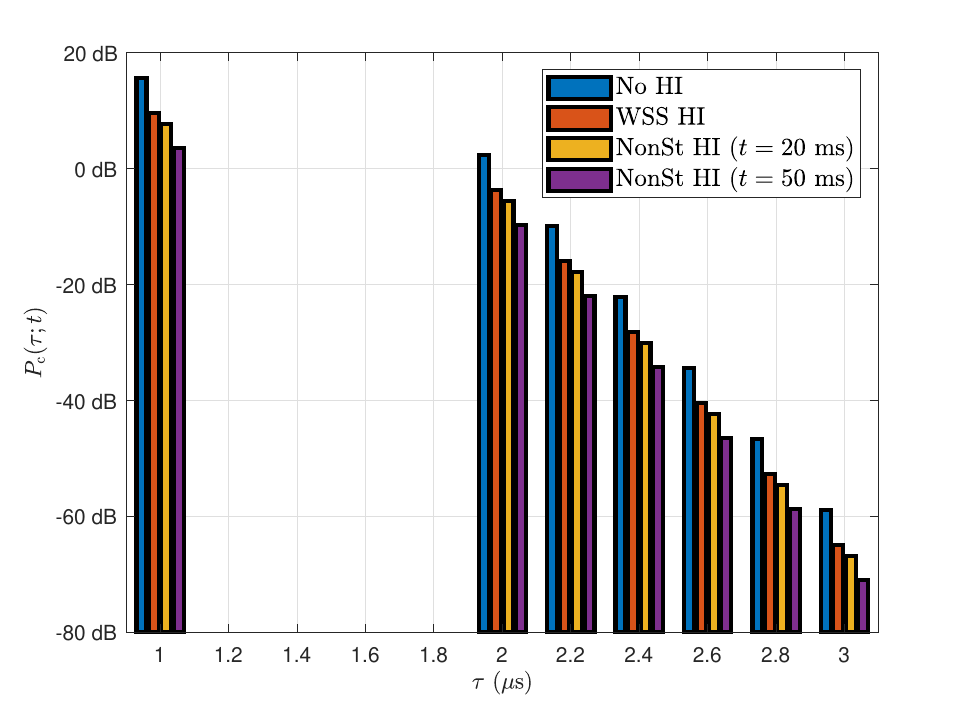}
		\vspace{-0.3cm}
		\caption{PDP for ideal and impaired hardware scenarios. Assumptions: $N = 20$, $K = 11.5$, and $\kappa_{\chi_{\rm T}}^2 = \kappa_{\chi_{\rm R}}^2 = 0.5$ W. HI stands for hardware impairment.}
		\vspace{-0.29cm}
		\label{fig:PDP}
	\end{minipage}\hfill
	\begin{minipage}{0.49\columnwidth}
		\centering
		\includegraphics[width=1\textwidth]{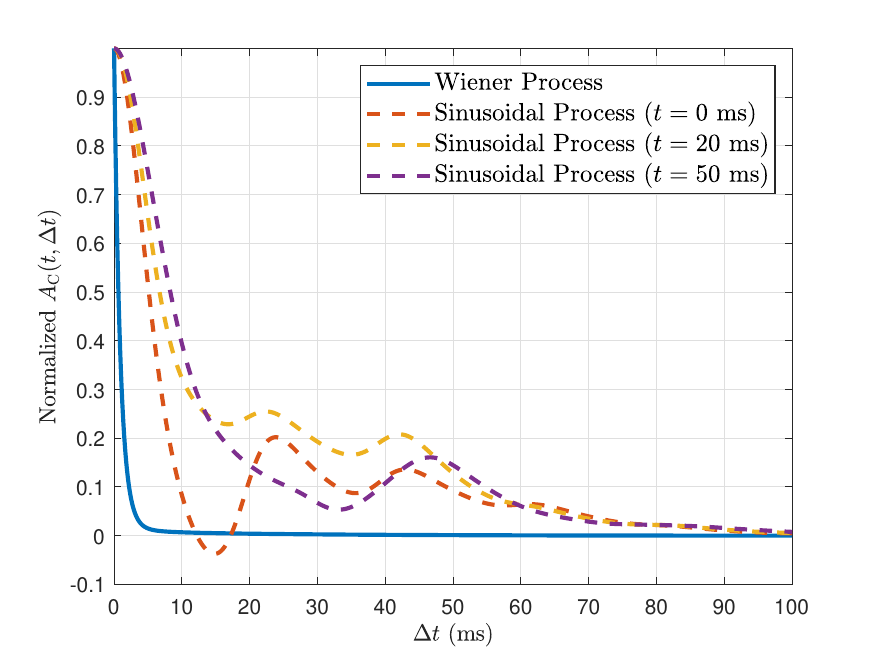}
		\vspace{-0.3cm}
		\caption{Coherence time: Normalized channel ACF for different random processes. Assumptions: $N = 20$, $K = 11.5$, $f_{\rm c} = 6$ GHz, $\theta_{\rm m} = 5^\circ$, and $l_{\chi_{\rm T}} = l_{\chi_{\rm R}} = 0.05$ s.}
		\vspace{-0.29cm}
		\label{fig:CT_1}
	\end{minipage}
\end{figure}

\subsection{PDP}
In Fig. \ref{fig:PDP}, we plot the PDP for different scenarios. In particular, we highlight the impact of hardware impairments on the PDP by comparing the ideal case (no hardware impairment) with those of impaired scenarios. The gap in the power of the received signal between the ideal and impaired cases is clear in this figure, which is around $6$ dB for WSS hardware impairments with $\kappa_{\chi_{\rm T}}^2 = \kappa_{\chi_{\rm R}}^2 = 0.5$ W. As shown before, note that UAV wobbling has no impact on the PDP. It can also be observed that the difference between the WSS and nonstationary hardware impairments in terms of their respective PDPs is not very high. This observation further motivates the use of WSS processes for modeling hardware impairments as they are more tractable than nonstationary models, which are otherwise more realistic \cite[Sec. 7.2.3]{B_Schenk_Imperfections_2008}.

\subsection{Coherence Time}
\begin{figure}[t!]
	\centering
	\begin{minipage}{0.49\columnwidth}
		\centering
		\includegraphics[width=1\textwidth]{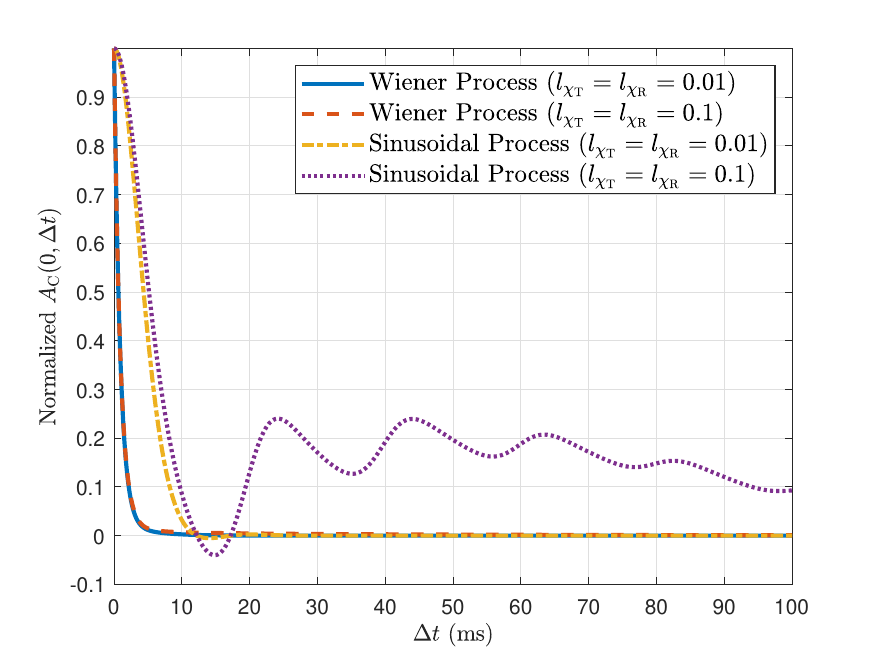}
		\vspace{-0.3cm}
		\caption{Coherence time: Normalized channel ACF for different random processes and impairment levels. Assumptions: $N = 20$, $K = 11.5$, $f_{\rm c} = 6$ GHz, $\theta_{\rm m} = 5^\circ$.}
		\vspace{-0.29cm}
		\label{fig:CT_2}
	\end{minipage}\hfill
	\begin{minipage}{0.49\columnwidth}
		\centering
		\includegraphics[width=1\textwidth]{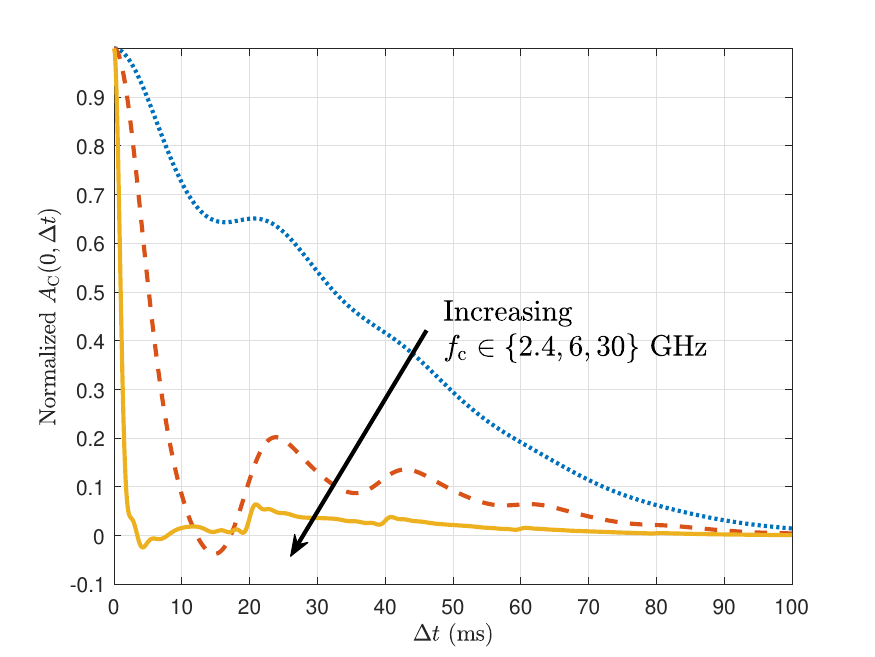}
		\vspace{-0.3cm}
		\caption{Coherence time: Normalized channel ACF for different $f_{\rm c}$. Assumptions: $K = 11.5$, $\theta_{\rm m} = 5^\circ$, and $l_{\chi_{\rm T}} = l_{\chi_{\rm R}} = 0.05$ s.}
		\vspace{-0.29cm}
		\label{fig:CT_3}
	\end{minipage}
\end{figure}

\begin{figure}[t!]
	\centering
	\begin{minipage}{0.49\columnwidth}
		\centering
		\includegraphics[width=1\textwidth]{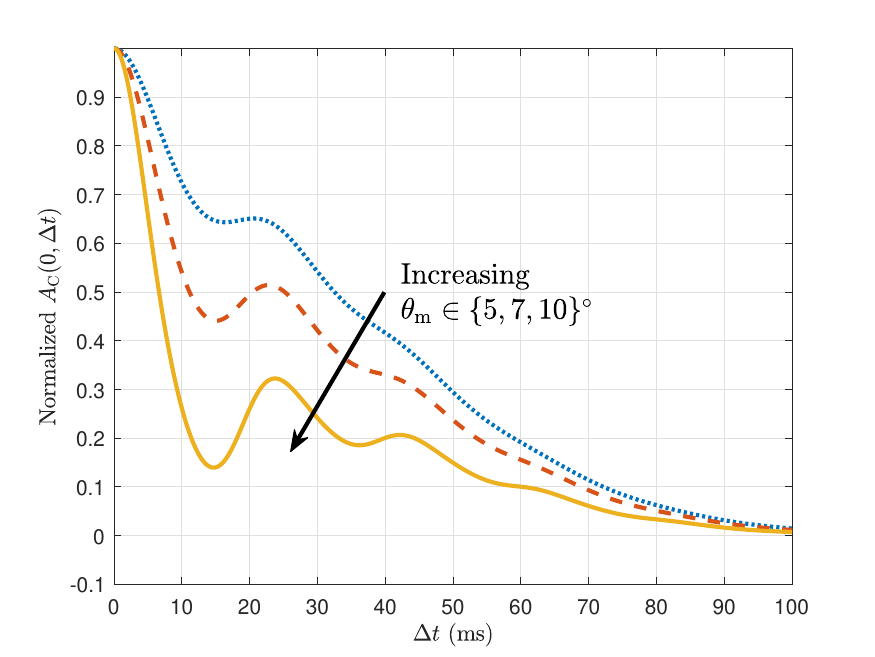}
		\vspace{-0.3cm}
		\caption{Coherence time: Normalized channel ACF for different $\theta_{\rm m}$. Assumptions: $K = 11.5$, $f_{\rm c} = 2.4$ GHz, and $l_{\chi_{\rm T}} = l_{\chi_{\rm R}} = 0.05$ s.}
		\vspace{-0.29cm}
		\label{fig:CT_4}
	\end{minipage}\hfill
	\begin{minipage}{0.49\columnwidth}
		\centering
		\includegraphics[width=1\textwidth]{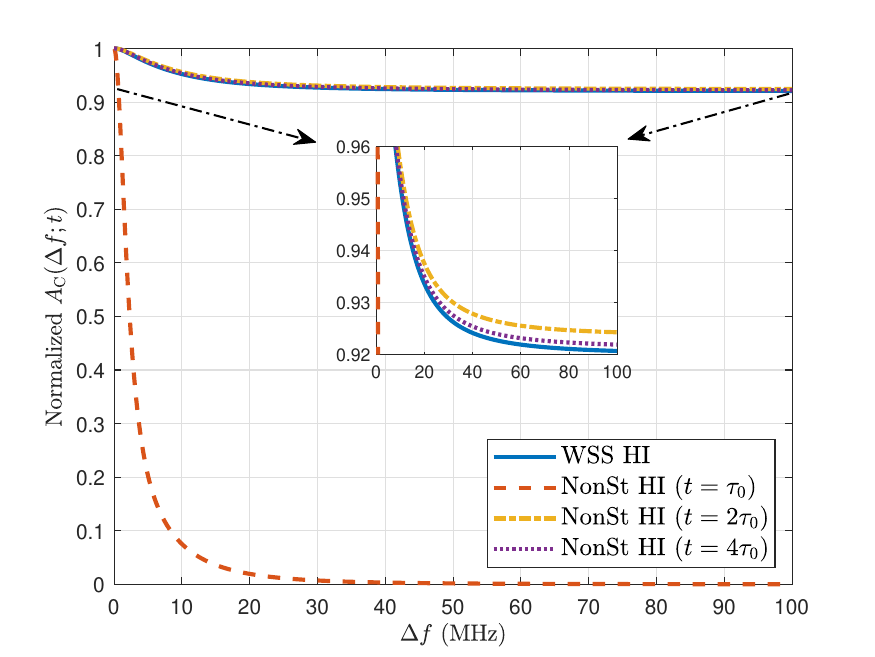}
		\vspace{-0.3cm}
		\caption{Coherence bandwidth: Normalized channel ACF for different impairment models and times $t$. Assumptions: $N = 20$ and $K = 11.5$.}
		\vspace{-0.29cm}
		\label{fig:CB_1}
	\end{minipage}
\end{figure}

We show the impact of UAV wobbling and hardware impairments on the channel coherence time in Figs. \ref{fig:CT_1}-\ref{fig:CT_4}. The general trends of the normalized channel ACF for the Wiener and sinusoidal processes are depicted in Fig. \ref{fig:CT_1}. Note that we used the WSS hardware impairment model to obtain these figures. Since the channel ACF becomes nonstationary for the sinusoidal process, we plotted it at different time instants to show its progression over time. Also, from this figure, we observe that channel coherence time is minimized at $t=0$ s, and thus, we use this time instant as a representative one to determine the channel coherence time. Assuming that $\gamma_{\rm T} = 0.5$, we find that $T_{\rm Coh}^{\rm WSS}(0) = 5.13$ ms and $T_{\rm Coh}^{\rm WSS-SI} = 643~\mu$s, which makes sense intuitively, as the variations in $\theta(t)$ for the Wiener process can grow without bounds. In Fig. \ref{fig:CT_2}, we compare the normalized channel ACF for different random processes and impairment levels. As seen in this figure, hardware impairments do not have any noticeable effect on the channel ACF when $\theta(t)$ follows the Wiener process. Although this was expected, we observe that hardware impairments do not severely degrade coherence time for the sinusoidal process either. In particular, we have $T_{\rm Coh}^{\rm WSS-SI} = \{639, 643\}~\mu$s and $T_{\rm Coh}^{\rm WSS}(0) = \{4.36, 5.16\}$ ms for $l_{\chi_{\rm T}} = l_{\chi_{\rm R}} = \{0.01, 0.1\}$ s. The impact of carrier frequency and maximum UAV pitch angle on the coherence time for the sinusoidal process is depicted in Figs. \ref{fig:CT_3} and \ref{fig:CT_4}. From a physical standpoint, we observe that the channel phase directly depends on the UAV-UE distance and the carrier frequency (see \eqref{phi_i}). Therefore, increasing $\theta_{\rm m}$ (thus, the UAV-UE distance) or $f_{\rm c}$ results in higher channel variations, which, in turn, decreases the channel coherence time. Numerically speaking, we have $T_{\rm Coh}^{\rm WSS}(0) = \{32.52, 5.13, 0.98\}$ ms for $f_{\rm c} = \{2.4, 6, 30\}$ GHz and $\theta_{\rm m} = 5^\circ$ and $T_{\rm Coh}^{\rm WSS}(0) = \{32.52, 11.13, 6.63\}$ ms for $\theta_{\rm m} = \{5, 7, 10\}^\circ$ and $f_{\rm c} = 2.4$ GHz. As is clear from these numbers, channel coherence time is heavily degraded at high frequencies due to UAV wobbling and hardware impairments (less than $1$ ms at $f_{\rm c} = 30$ GHz), which renders channel estimation and tracking very challenging.

\begin{figure}[t!]
	\centering
	\begin{minipage}{0.49\columnwidth}
		\centering
		\includegraphics[width=1\textwidth]{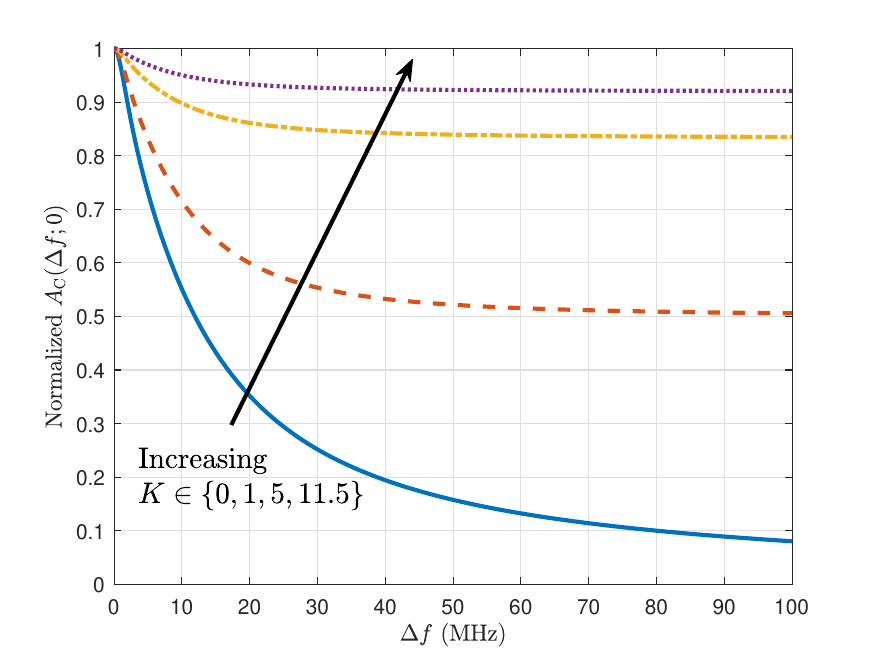}
		\vspace{-0.3cm}
		\caption{Coherence bandwidth: Normalized channel ACF for different Rician $K$ factors. Assumptions: WSS hardware impairments and $N = 20$.}
		\vspace{-0.29cm}
		\label{fig:CB_2}
	\end{minipage}\hfill
	\begin{minipage}{0.49\columnwidth}
		\centering
		\includegraphics[width=1\textwidth]{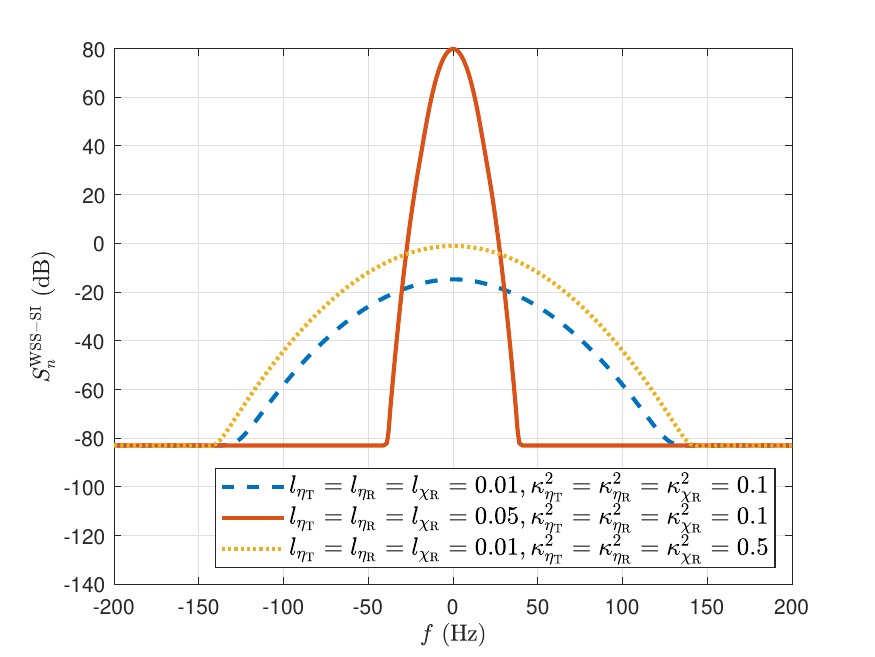}
		\vspace{-0.3cm}
		\caption{PSD of $n(t)$ for different impairment levels. Assumptions: Wiener wobbling, WSS hardware impairments, $N = 20$, $K = 11.5$, $f_{\rm c} = 2.4$ GHz, and $\theta_{\rm m} = 5^\circ$.}
		\vspace{-0.29cm}
		\label{fig:PSD_1}
	\end{minipage}
\end{figure}

\subsection{Coherence Bandwidth}
In Figs. \ref{fig:CB_1} and \ref{fig:CB_2}, we plot the normalized channel ACF as a function of $\Delta f$ to investigate the impact of different impairments on the channel coherence bandwidth. From \eqref{B_Coh_WSS}, we observe for WSS hardware impairments that the normalized ACF monotonically goes from $1$ to $\frac{K}{K+1}$ as $\Delta f$ goes from $0$ to $\infty$. Interestingly, Fig. \ref{fig:CB_1} shows that this is also the case for nonstationary hardware impairments as long as $t > \tau_0$. Assuming that $\gamma_{\rm B} = 0.95$, we have $B_{\rm Coh}^{\rm WSS} = 10.75$ MHz and $B_{\rm Coh}^{\rm NonSt}(t) = \{0.51, 12.21, 11.21\}$ MHz at $t = \{\tau_0, 2\tau_0, 4\tau_0\}$. Note that the behavior of the nonstationary hardware impairments quickly converges to that of the WSS one. In Fig. \ref{fig:CB_2}, we show the channel ACF at time instant $t = 0$ s for different Rician $K$ factors. As the power of the LoS link increases with respect to the powers of other MPCs, we expect to have lower delay spreads, and thus, higher coherence bandwidths.

\subsection{PSD of Distortion-Plus-Noise}
\begin{figure}[t!]
	\centering
	\begin{minipage}{0.49\columnwidth}
		\centering
		\includegraphics[width=1\textwidth]{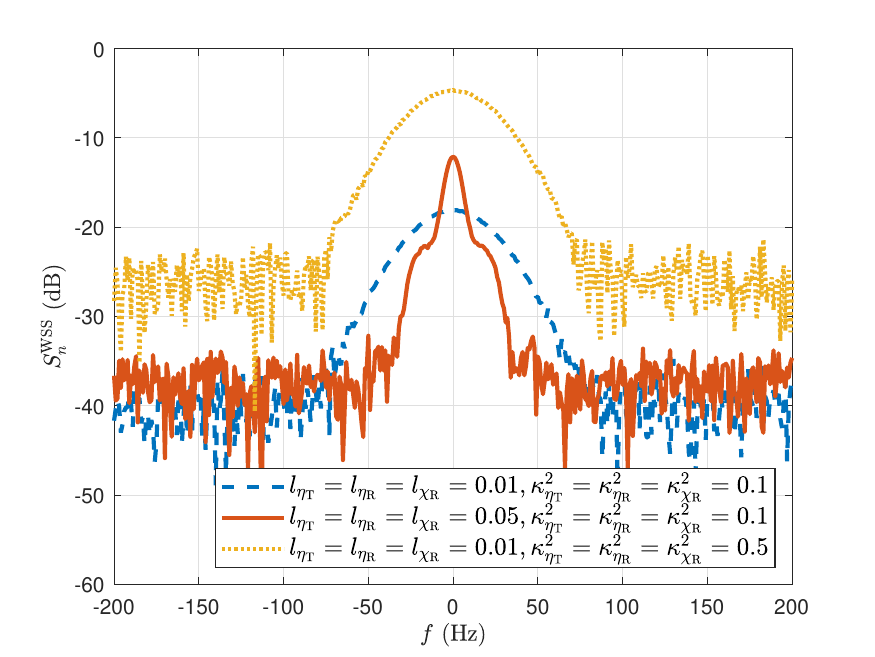}
		\vspace{-0.3cm}
		\caption{PSD of $n(t)$ for different impairment levels. Assumptions: Sinusoidal wobbling, WSS hardware impairments, $N = 20$, $K = 11.5$, $f_{\rm c} = 2.4$ GHz, $\theta_{\rm m} = 5^\circ$.}
		\vspace{-0.29cm}
		\label{fig:PSD_2}
	\end{minipage}\hfill
	\begin{minipage}{0.49\columnwidth}
		\centering
		\includegraphics[width=1\textwidth]{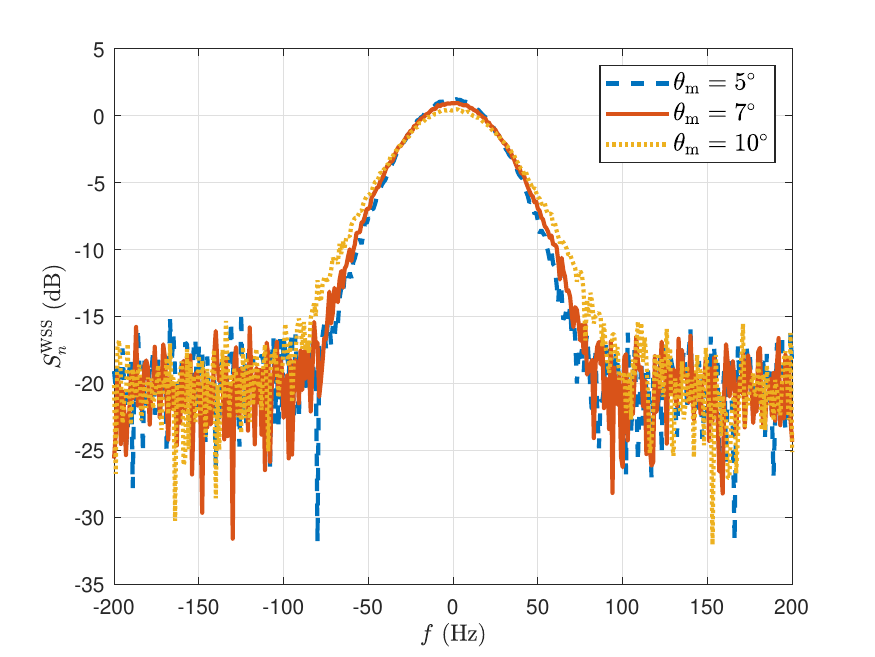}
		\vspace{-0.3cm}
		\caption{PSD of $n(t)$ for different $\theta_{\rm m}$. Assumptions: WSS hardware impairments, $N = 20$, $K = 11.5$, $f_{\rm c} = 2.4$ GHz, $\kappa_{\eta_{\rm T}}^2 = \kappa_{\eta_{\rm R}}^2 = \kappa_{\chi_{\rm R}}^2 = 1$ W, $l_{\eta_{\rm T}} \!=\! l_{\eta_{\rm R}} \!=\! l_{\chi_{\rm R}} \!=\! 0.01\!$ s.}
		\vspace{-0.29cm}
		\label{fig:PSD_3}
	\end{minipage}
\end{figure}

We provide numerical results for the PSD of $n(t)$ in Figs. \ref{fig:PSD_1}-\ref{fig:PSD_3}. Assuming both Wiener and sinusoidal models for UAV wobbling, we plot the PSD of $n(t)$ for different hardware impairment levels in Figs. \ref{fig:PSD_1} and \ref{fig:PSD_2}, respectively. We observe that increasing the maximum power of the WSS hardware impairment model ($\kappa^2$ parameters, e.g., $\kappa_{\eta_{\rm T}}^2$) while keeping the characteristic length-scale of this model ($l$ parameters, e.g., $l_{\eta_{\rm T}}$) constant simply shifts the PSD upward (in dB), which means that the total power of $n(t)$ (integral of its PSD) increases. On the other hand, increasing the $l$ parameters while keeping the $\kappa^2$ parameters constant squeezes the PSD curve toward the $0$ Hz frequency, which could either increase or decrease the total distortion-plus-noise power depending on the specific values of the $l$ and $\kappa^2$ parameters and the wobbling model. In Fig. \ref{fig:PSD_3}, we assume that UAV wobbling follows a sinusoidal model and compare the PSD of $n(t)$ for different values of the maximum UAV pitch angle. As is clear in this figure and assuming realistic values for $\theta_{\rm m}$, the PSD of $n(t)$ is resilient to this impairment and does not degrade noticeably with increasing the maximum UAV pitch angle.

\begin{remark}
Hardware impairments are the primary factors in determining the PDP, coherence bandwidth of the channel, and PSD of distortion-plus-noise process, while UAV wobbling dominates the coherence time of the channel.
\end{remark}

\section{Conclusion} \label{sec:conclusion}
In this paper, we developed a unified air-to-ground channel model that accounts for multiple impairment types. Specifically, we considered a multi-path Rician channel, where a ground UE receives data from a hovering UAV via one LoS and multiple NLoS paths. Two fundamentally different impairment types have been considered in this paper, i.e., UAV wobbling and hardware impairments (RF nonidealities at both the UAV and the UE). We modeled the UAV pitch angle wobbling by two different stochastic processes, one with independent and stationary increments (the Wiener process) and another with bounded fluctuations to represent the oscillatory nature of wobbling (the sinusoidal process). As for the RF nonidealities, we modeled the combined effect of all hardware impairments as two multiplicative and additive distortion noise processes. Using these models, we concretely analyzed the channel ACF and provided easy-to-use equations for four key channel-related metrics, i.e., PDP, coherence time, coherence bandwidth, and PSD of the distortion-plus-noise process. Our results demonstrated a heavy degradation of the channel coherence time at high carrier frequencies due to UAV wobbling and hardware impairments, which effectively makes channel estimation more challenging at mmWave and higher frequencies. To the best of our knowledge, this is the first work that provides a comprehensive analysis of the joint effect of UAV wobbling and hardware impairments on the air-to-ground wireless channel. One meaningful extension of this work is to provide the same analysis when the UAV is mobile, which makes the effective Doppler phase shift even more complicated. Also, analyzing the air-to-ground impairments-aware wireless channel when centralized or distributed MIMO setups are considered (instead of single-antenna transceivers) is another future line of research.

\vspace{-0.2cm}
\bibliographystyle{IEEEtran}
\bibliography{../../AllReferences}

\end{document}